\documentclass[universe,article,accept,pdftex,moreauthors]{Definitions/mdpi}

\firstpage{1} 
\pubvolume{1}
\issuenum{1}
\articlenumber{0}
\pubyear{2022}
\copyrightyear{2022}
\externaleditor{{Academic Editor:} 
 }
\datereceived{01 June 2022}
\dateaccepted{28 June 2022} 
\datepublished{} 
\hreflink{https://doi.org/} 

\Title{Incorporating a Radiative Hydrodynamics Scheme in the Numerical-Relativity Code BAM}
\TitleCitation{Incorporating a Radiative Hydrodynamics Scheme in the Numerical-Relativity Code BAM}

\Author{Henrique Gieg 
 $^{1,}$*, Federico Schianchi $^{2}$, Tim Dietrich $^{2,3}$ and Maximiliano Ujevic $^{1}$}

\AuthorNames{Henrique Gieg, Federico Schianchi, Tim Dietrich and Maximiliano Ujevic}

\AuthorCitation{Gieg, H.; Schianchi, F.; Dietrich, T.; Ujevic, M.}

\address{${}^1$ \quad Centro de Ci\^encias Naturais e Humanas, Universidade Federal do ABC, Santo Andr\'e 09210-170, SP, Brazil; mujevic@ufabc.edu.br \\
${}^2$ \quad Institut f\"ur Physik und Astronomie, Universit\"at Potsdam, Haus 28, Karl-Liebknecht-Str. 24/25, \mbox{14476 Potsdam, Germany}; schianchi@uni-potsdam.de (F.S.); tim.dietrich@uni-potsdam.de (T.D.) \\
${}^3$ \quad Max Planck Institute for Gravitational Physics (Albert Einstein Institute), Am M\"uhlenberg 1, \mbox{14476 Potsdam, Germany}}

\corres{Correspondence: henrique@gieg.com.br}

\abstract{
\textls[-20]{To study binary neutron star systems and to interpret observational data such as} gravitational-wave and kilonova signals, one needs an accurate description of the processes that take place during the final stages of the coalescence, for example, through numerical-relativity simulations. In this work, we present an updated version of the numerical-relativity code BAM in order to incorporate nuclear-theory-based equations of state and a simple description of neutrino interactions through a neutrino leakage scheme. Different test simulations, for stars undergoing a neutrino-induced gravitational collapse and for binary neutron stars systems, validate our new implementation. For the binary neutron stars systems, we show that we can evolve stably and accurately distinct microphysical models employing the different equations of state: SFHo, DD2, and the hyperonic BHB$\Lambda \phi$. Overall, our test simulations have good agreement with those reported in the literature.} 


\begin{document}


\section{Introduction}
In August 2017, the Advanced LIGO~\cite{LIGOScientific:2014pky} and Advanced Virgo~\cite{VIRGO:2014yos} gravitational wave (GW) interferometers detected, for the first time, a GW signal arising from the merger of two neutron stars (NSs) (GW170817)~\cite{TheLIGOScientific:2017qsa}. This GW detection was accompanied by a variety of electromagnetic (EM) signatures across the entire frequency spectrum~\cite{LIGOScientific:2017ync}. The observed signals, GWs and EM, were created by a binary neutron star (BNS) merger that happened about $130$ million years ago in the galaxy NGC 4993~\cite{LIGOScientific:2017ync}. 
While the GW signal was emitted during the inspiral of the stars before the merger, the EM signals were created after the merger. These include the short gamma-ray burst GRB170817A~\cite{LIGOScientific:2017ync} observed 1.7 s after the stars' collision, the weeks-long kilonova AT2017gfo~\cite{Arcavi:2017xiz,Coulter:2017wya,Lipunov:2017dwd,DES:2017kbs,Tanvir:2017pws,Valenti:2017ngx}, and sGRB/kilonova afterglows that are still observable~\cite{Hajela:2019mjy,Hajela:2021faz}.

Over the last years, this landmark discovery has been extensively studied, yielding constraints not only on the NS properties, such as radius, tidal deformability, and its equation of state (EoS)~\cite{Annala:2017llu,Bauswein:2017vtn,Fattoyev:2017jql,Ruiz:2017due,Shibata:2017xdx,Radice:2017lry,Most:2018hfd,Tews:2018iwm,Coughlin:2018miv,Coughlin:2018fis,Capano:2019eae,Dietrich:2020efo,Nedora:2021eoj,Huth:2021bsp}, but also on the expansion rate of our universe~\cite{LIGOScientific:2017adf,Guidorzi:2017ogy,Hotokezaka:2018dfi,Coughlin:2019vtv,Dietrich:2020efo,Perez-Garcia:2022gcg,Wang:2020vgr,Bulla:2022ppy}. However, we still have not fully understood the internal NS structure, the composition, and the underlying physics since many modeling aspects are plagued by large uncertainties.

Indeed, for a correct interpretation of the observables, one has to correlate the observational data with theoretical predictions. For the development of such models, numerical-relativity (NR) simulations are an important prerequisite as they provide a testbed for new GW models (e.g.,~\cite{Dietrich:2020eud} and references therein), and they enable us to connect properties of the outflowing matter to the binary properties~\cite{Dietrich:2016fpt,Coughlin:2018miv,Radice:2018pdn,Coughlin:2018fis,Nedora:2020qtd,Dietrich:2020efo,Kruger:2020gig}. However, to achieve this, we need, among other things, (i) a meticulous treatment of the stellar matter with state-of-art nuclear theory EoSs to account for temperature and composition dynamics and (ii) an approach to incorporate microphysical processes such as neutrino-driven reactions that {are} related to nucleosynthesis. In particular, requirement (ii) is corroborated by the observed kilonova AT2017gfo, which suggests the importance of $r$-process nucleosynthesis~\cite{Metzger:2010sy,Watson:2019xjv} in neutron star merger outflows. 

In this work, we explain recent updates to the NR code BAM~\cite{Bruegmann:2006at,Thierfelder:2011yi,Dietrich:2015iva,Bernuzzi:2016pie} {(bifunctional adaptative mesh)}, focusing on the implementation of a neutrino leakage scheme \linebreak  (NLS)~\mbox{\cite{Ruffert:1995fs, Rosswog:2003rv}} to describe neutrino production and transport using tabulated nuclear-theory based EoSs, and subsequent modifications to the general relativistic hydrodynamics (GRHD) routines. We validate our code extensions with a variety of tests and present a set of new BNS simulations. 

The structure of this article is as follows. In Section~\ref{sec:2}, we discuss the underlying theory and summarize the basic equations that we modified in order to incorporate neutrino interactions. In Section~\ref{sec:3}, we outline the employed numerical methods and implementation. Tests for our new scheme---that is, single isolated neutron star {(modeled as solutions of the Tolman--Oppenheimer--Volkoff~\cite{Tolman:1939jz} (TOV) structure equations)} evolutions undergoing neutrino-induced collapse---are shown in Section~\ref{sec:4}. In Section~\ref{sec:5}, we present BNS simulations, and we conclude in Section~\ref{sec:6}. Throughout this work, we employ geometric units ($c = G = 1$), and we set the Boltzmann constant and the mass of the sun equal to one, that is, $M_\odot = k_B = 1$. The metric signature is $(-,+,+,+)$; Greek indices $\mu, \nu, \dots$ run from $0$--$3$, while Latin indices $i, j, \dots$ run from $1$--$3$; and Einstein's summation convention is employed.

\section{\label{sec:2}Fundamental Equations}

\subsection{3 + 1 Decomposition and Spacetime Evolution}


BAM employs the {(3 + 1)-dimensional Arnowitt--Deser--Misner (ADM)} decomposition formalism, that is, the four-dimensional spacetime is foliated by a set of nonintersecting \mbox{three-dimensional} spacelike hypersurfaces $\Sigma$, with a field of timelike normal vectors $n^\mu$. The spacetime coordinates are chosen such that the line element reads
\begin{equation}\label{eq:3+1-line}
    ds^2 = -\alpha^2 dt^2 + \gamma_{ij}(\beta^i dt + dx^i)(\beta^j dt+ dx^j),
\end{equation}
where $\alpha>0$ is the lapse function, $\gamma_{\mu \nu} = g_{\mu \nu} + n_\mu n_\nu$ is the spatial metric induced on the hypersurfaces $\Sigma$, $\beta^i$ is the spatial shift vector, and $dx^i$ is the spatial coordinates displacement.
Likewise, the components of the normal field are given by
\begin{eqnarray}
n^\mu &=& (\alpha^{-1}, -\alpha^{-1}\beta^i), \\
n_\mu &=& (-\alpha, 0, 0, 0).
\end{eqnarray}

For the dynamical evolution of the spacetime, we are using the BSSNOK scheme~(\cite{Baumgarte:1998te} and references therein) for the TOV runs {in order to compare our results with those reported in the literature}, while we make use of the Z4c scheme~for the BNS runs (\cite{Hilditch:2012fp} and references therein) {because of its constraint violation damping properties, which allow more accurate solutions of the Einstein field equations, especially in the presence of matter.}

\subsection{General Relativistic Radiative Hydrodynamics}
The covariant GRHD equations arise from the relevant conservation laws. The first of these is the baryon number conservation
\begin{equation}\label{eq:bar-cons-cov}
    \nabla_\mu(\rho u^\mu) = 0,
\end{equation}
where $\nabla_\mu$ is the covariant derivative compatible with the spacetime metric $g_{\mu \nu}$; $\rho = m_b n_b$ is the rest-mass density, with $m_b$ being a baryon mass constant chosen depending on the EoS and $n_b$ being the baryon number density; and $u^\mu$ is the matter element four-velocity, which in terms of 3+1 fields is written as

\begin{equation}
    u^\mu = W(n^\mu + v^\mu),
\end{equation}
where $v^\mu$ is the spatial velocity measured by the Eulerian frame $v^\mu n_\mu = 0$; $W = 1/\sqrt{1-v^2}$ is the Lorentz factor; and $v^2 = \gamma_{ij}v^iv^j$.

The second equation is the energy-momentum conservation, given by
\begin{equation}\label{eq:s-t-cons-tot}
    \nabla_\nu T^{\mu \nu} = 0,
\end{equation}
where $T_{\mu \nu}$ is the stress-energy tensor. Within this work, we describe matter as an ideal \mbox{fluid, hence}
\begin{equation}
     T^{\mu \nu}_{\rm fluid} = (e + p)u^\mu u^\nu + pg^{\mu \nu},
\end{equation}
where $e$ is the energy density, and $p$ is the pressure measured in the fluid comoving frame. If the spacetime is filled with matter and neutrinos, the stress-energy tensor becomes
\begin{equation}\label{eq:tot-s-e}
    T^{\mu \nu} = T^{\mu \nu}_{\rm fluid} + T^{\mu \nu}_{\rm rad},
\end{equation}
where $T^{\mu \nu}_{\rm rad}$ is the stress-energy tensor of the neutrinos, hereby modeled as radiation. Therefore,  Equation~\eqref{eq:s-t-cons-tot} reads
\begin{equation}\label{eq:s-t-cons-fl}
    \nabla_\nu T^{\mu \nu}_{\rm fluid} = - \nabla_\nu T^{\mu \nu}_{\rm rad} \equiv \Psi^\mu,
\end{equation}
where we defined for convenience $\Psi^\mu \equiv - \nabla_\nu T^{\mu \nu}_{\rm rad}$. Equation~\eqref{eq:s-t-cons-fl} then states that energy and momentum are carried away by neutrinos, producing variations on the energy and momentum of a fluid element. 

In addition, incorporating neutrino-driven reactions, the conservation of leptons must be enforced explicitly. {For simplicity, we assume that the only lepton species in} the fluid {are} electrons and positrons. {Hence,} the relevant conservation law reads
\begin{equation}\label{eq:lep-cons-cov}
    \nabla_\mu(\rho Y_e u^\mu) = \rho \mathcal{R},
\end{equation}
where $Y_e = n_e/n_b$ is the electron fraction, {$n_e = n_{e^{-}} - n_{e^{+}}$ is the net electron number density, $n_{e^{-}}$ is the number density of electrons, $n_{e^{+}}$ is the number density of positrons,} and $\mathcal{R}$ is a source term accounting for the variations of the lepton number within a matter element in response to the emission/absorption of neutrinos. In fact, Equation~\eqref{eq:lep-cons-cov} implies, with the help of Equation~(\ref{eq:bar-cons-cov}), that
\begin{equation}
    u^\mu \nabla_\mu Y_e = \frac{d Y_e}{d\tau}= \mathcal{R},
\end{equation}
where $d\tau$ is the proper time elapsed for a matter element. Hence, $\mathcal{R}$ can be understood as the rate of change of the electron fraction measured in the fluid rest-frame.

The next step is to bring Equations~\eqref{eq:bar-cons-cov},~\eqref{eq:s-t-cons-fl}, and~\eqref{eq:lep-cons-cov}, which constitute the GRHD equations in covariant formulation, into the correspondent coordinate expressions as the following balance law: 
\begin{equation}\label{eq:grhd-val-form}
\partial_0\textbf{q} + \partial_i \textbf{F}^{i}(\textbf{q}) = \textbf{S}(\textbf{q}),
\end{equation}
known as the Valencia formulation of the GRHD equations~\cite{Banyuls:1997zz}. Equation~\eqref{eq:grhd-val-form} is a system of six partial differential equations that performs the time evolution of the \textit{conserved}
quantities
\begin{equation}
\textbf{q} \equiv \sqrt{\gamma}\left(\begin{array}{c} D \\ \tau \\ S_k \\ DY_e \end{array}\right) = \sqrt{\gamma}\left(\begin{array}{c} W\rho \\ \rho h W^2 - p - W\rho \\ \rho h W^2 v_k \\ \rho W Y_e \end{array}\right),
\end{equation}
where $\gamma$ is the determinant of the spatial metric, $D$ is the rest-mass density, $\tau$ is the energy density, $S_k$ is the momentum density, and $DY_e$ is the conserved electron fraction, all measured in the Eulerian frame, which are defined in terms of the \textit{primitive} quantities 

\begin{equation}
\textbf{w} \equiv (\rho, \epsilon, v_k, p, Y_e),
\end{equation}
where $\epsilon$ is the specific internal energy per baryon, and $h = 1 + \epsilon + p/\rho$ is the specific enthalpy per baryon. The fluxes are
\begin{eqnarray}\label{eq:flx}
\textbf{F}^i \equiv \sqrt{\gamma}\left(\begin{array}{c} D(\alpha v^i - \beta^i) \\ \tau(\alpha v^i - \beta^i) + \alpha p v^i \\
S_k(\alpha v^i - \beta^i) + \alpha p \delta^i{_k}\\
DY_e(\alpha v^i - \beta^i)\end{array}\right),
\end{eqnarray}
and the source terms read
\begin{equation}\label{eq:sources}
\textbf{S} \equiv \sqrt{\gamma}\left(\begin{array}{c} 0 \\ \alpha \mathcal{S}^{ij}K_{ij} - S^i\partial_i \alpha + \alpha^2 \Psi^0 \\
\frac{\alpha}{2}\mathcal{S}^{ij}\partial_k\gamma_{ij} + S_i \partial_k \beta^i -(\tau + D) \partial_k \alpha + \alpha \Psi_k \\
\alpha \rho \mathcal{R}
\end{array}\right),
\end{equation}
with $\mathcal{S}_{ij} \equiv \gamma_{i\mu}\gamma_{j \nu} T^{\mu \nu}_{\rm fluid}$ being the spatial stress tensor of the matter distribution.

In order to close the GRHD system of equations, an EoS must be provided to compute the pressure $p$ from the remaining primitives. One of our new additions to the BAM code is related to this point. Instead of providing as input a one-dimensional EoS $p_{\rm cold}(\rho)$ parametrized as a piecewise polytrope~\cite{Read:2008iy} augmented with a $\Gamma$-law EoS to model thermal effects (i.e., $p = p_{\rm cold} + p_{\rm th}$ with $p_{\rm th} = (\Gamma - 1)\rho\epsilon$~\cite{Shibata:2005ss}), we consider more general and realistic nuclear-theory EoSs in the form of three-dimensional tables. In this scenario, the necessary thermodynamical quantities are represented as functions of the rest-mass density, temperature, and electron fraction, and are computed via trilinear interpolations.




\subsection{Neutrino Leakage}

The NLS has been employed for a variety of astrophysical systems to model neutrino emission (e.g., core-collapse supernovae~\cite{1985ApJS...58..771B, 2010CQGra..27k4103O, OConnor:2014sgn} and compact binary mergers~\mbox{\cite{Ruffert:1995fs, Rosswog:2003rv, Deaton:2013sla, Foucart:2014nda, Foucart:2015gaa, Neilsen:2014hha}}). It possesses a number of advantages, such as (i) a simple implementation; (ii) reasonable (qualitative) description of neutrinos' features in NSs, particularly in optically thick media~(\cite{Foucart:2015gaa} and references therein); and (iii) low computational costs. Therefore, its implementation in numerical-relativity simulations is compelling. Moreover, despite the underlying assumptions of the approach (which will become clear in the following), the NLS serves as a first approximation for radiative losses and is a basis to support more intricate and realistic methods, for example, in radiation transport moment schemes~\cite{Shibata:2011kx, Cardall:2013kwa, Foucart:2015vpa, Anninos:2020bpo, Foucart:2020qjb, Weih:2020wpo, Foucart:2021mcb, Radice:2021jtw}, Lattice-Boltzmann methods~\cite{Weih:2020qyh}, leakage-equilibration-absorption schemes~\cite{Ardevol-Pulpillo:2018btx}, and advanced leakage schemes~\cite{Perego:2015agy,Gizzi:2019awu,Gizzi:2021ssk}.

\subsubsection*{Underlying Hypotheses of the Neutrino Leakage Scheme}

The NLS is characterized by a number of hypotheses or assumptions (outlined in, e.g., \cite{Galeazzi:2013mia}). For completeness, we will describe the most important aspects in \mbox{the following}: 

\begin{enumerate}[leftmargin=*,labelsep=5mm]
\item {For 
 simplicity, we consider only} electrons and positrons {as representative leptons within the fluid}. 
\item The considered neutrino flavors are electron neutrinos $\nu_e$, electron antineutrinos $\bar\nu_e$, and heavy lepton neutrinos/antineutrinos $\nu_{\mu, \tau},~\bar\nu_{\mu, \tau}$, collectively grouped as a single species $\nu_x$ with statistical weight~4. 
\item Neutrinos obey the ultra-relativistic Fermi--Dirac distribution in local $\beta$-equilibrium and have the same temperature as the matter. Hence, the relativistic chemical potentials (i.e., including rest-masses of protons, neutrons, and electrons) for electron-flavored neutrinos read
\end{enumerate}

\begin{equation}\label{eq:beta-eq}
    \mu_{\nu_e} = -\mu_{\bar\nu_e} = \mu_p + \mu_e - \mu_n,
\end{equation}
where, for simplicity, we assume $\mu_{\nu_x} = 0$, given that heavy lepton neutrinos rarely interact with matter.
{This hypothesis is justified by the assumption that a possible non-equilibrium condition (induced, for instance, by a density oscillation) is rapidly driven to $\beta$-equilibrium on a timescale that is much smaller than the timestep adopted to numerically evolve the matter and spacetime quantities. However, it is important to point out that reestablishing $\beta$-equilibrium from a (short-lived) non-equilibrium state implies energy dissipation, which translates into damping of density oscillations by neutrinos bulk-viscosity~\cite{Alford:2019qtm, Alford:2019kdw}.}

{In the context of BNS mergers, \cite{Most:2021zvc} suggests that imprints of this effect in the late inspiral GW are undetectable, while during merger and post-merger, the bulk-viscosity may represent a non-negligible contribution to the damping. It is reported in \cite{Alford:2020lla} that in typical post-merger conditions, such a bulk viscous damping would operate at density oscillation frequencies $\lesssim 10~{\rm kHz}$ on a timescale of a few ms in the densest portions of the remnant. Therefore, this effect could impact the post-merger evolution within a simulation timespan, in particular the properties of ejecta with $T \lesssim 5~{\rm MeV}$. Finally, we remark that in the simplified approach of this work, bulk-viscosity is neglected, and further investigations of this topic are reserved for future works.}

\begin{enumerate}[leftmargin=*,labelsep=5mm]
\item[4.] The emission of neutrinos is isotropic in the fluid rest-frame and is given by
\begin{equation}\label{eq:psi-mu}
    \Psi^\mu = - n_b \mathcal{Q} u^\mu,
\end{equation}
where the total emissivity $\mathcal{Q}$ (energy per unit time and baryon) is the sum of emissivities for all neutrino flavors
\begin{equation}\label{eq:Q_tot}
    \mathcal{Q} \equiv Q(\nu_e) + Q(\bar\nu_e) + Q(\nu_x).
\end{equation}
To see that Equation~\eqref{eq:psi-mu} corresponds to an isotropic emission, note that the projection of $\Psi^\mu$ onto the hypersurface orthogonal to the fluid worldlines via the projector $h_{\mu \nu} = g_{\mu \nu} + u_{\mu}u_{\nu}$ vanishes (i.e., $h_{\mu \nu}\Psi^\mu = 0$). Hence, neutrinos are emitted such that no net momentum flux is perceived in the fluid comoving frame.

\item[5.] The source term $\mathcal{R}$ is given by
\begin{equation}\label{eq:R_tot}
    \mathcal{R} \equiv R(\bar\nu_e) - R(\nu_e),
\end{equation}
{where $R(\bar\nu_e)$ is the electron antineutrinos production rate, and $R(\nu_e)$ is the electron neutrinos production rate. Then, the Equation above states that} the creation of electron (anti-) neutrinos demand the (creation) annihilation of an (electron) positron in order to conserve the lepton family number.

\item[6.]Neutrinos are treated as a `test' fluid. Hence, the projections of $T^{\mu \nu}_{\rm rad}$, which act as sources of spacetime curvature, are neglected.
\end{enumerate}

Furthermore, the pressure and specific internal energy of a volume containing matter and neutrinos is given by
\begin{eqnarray}
 p = p_{\rm fluid} + p_{\nu_e} + p_{\bar\nu_e} + p_{\nu_x}\ , \hspace{1cm} \epsilon = \epsilon_{\rm fluid} + \epsilon_{\nu_e} +\epsilon_{\bar\nu_e} + \epsilon_{\nu_x}.\label{eq:eps-tot}
\end{eqnarray}
\indent Nevertheless, 
 the neutrino contributions to the above equations are only reasonable in opaque media (in which the neutrinos are said to be \textit{trapped}
), where radiation mostly diffuses in equilibrium with its surroundings. In semi-transparent media, where neutrinos rarely interact with matter, radiation flows as \textit{freely~streaming}; hence, no pressure is exerted by neutrinos, and no energy transfer occurs between matter and radiation. Besides, the spatial identification of trapped, freely streaming, and `gray' regimes within an NS is hardly feasible beforehand, and it is very difficult to capture and encode them in Equation~\eqref{eq:eps-tot}, at least in an NLS framework. Therefore, we opt for a simpler approach, in which the pressure and specific internal energy contributions of neutrinos are neglected within the whole extent of an NS.
It is straightforward to verify that if neutrinos are described by an ultra-relativistic Fermi--Dirac distribution, their pressure and specific internal energy contributions are only sizable in low-density, high-temperature regions, where interactions rarely occur. Thus, the error made in the approximation $p \approx p_{\rm fluid},~\epsilon \approx \epsilon_{\rm fluid}$ is negligible.



It is worth pointing out that the adoption of the `test' fluid hypothesis only simplifies our treatment of neutrinos with respect to their \textit{direct} role in the spacetime and matter evolutions. Thus, what remains is the way in which neutrinos alter the hydrodynamics (as in Equations~\eqref{eq:tau-evo}--\eqref{eq:DY-evo} below).

We end this section by explicitly showing how the previously introduced GRHD equations have to be modified following the previous hypotheses. While the baryon number conservation remains unaltered, the energy density, momentum density, and conserved electron fraction evolve, respectively, according to 
\vspace{-15pt}

\begin{adjustwidth}{-\extralength}{0cm}
\begin{eqnarray}
\partial_0(\sqrt{\gamma}\tau) + \partial_i[\sqrt{\gamma}\tau(\alpha v^i - \beta^i) + \sqrt{\gamma}\alpha p v^i] &=& \sqrt{\gamma}(\alpha \mathcal{S}^{ij}K_{ij} - S^i\partial_i\alpha) {- \alpha \sqrt{\gamma}\mathcal{Q} m_b^{-1} D}, \label{eq:tau-evo} \\
\partial_0(\sqrt{\gamma} S_k) + \partial_i[\sqrt{\gamma}S_k(\alpha v^i - \beta^i) + \sqrt{\gamma}\alpha p\delta^{i}{_k}] &=& \sqrt{\gamma}\left(\frac{\alpha}{2}\mathcal{S}^{ij}\partial_k\gamma_{ij} + S_i \partial_k \beta^i -(\tau + D)\partial_k \alpha\right)    {-\alpha\sqrt{\gamma}\mathcal{Q}m_b^{-1}D v_k}, \\
\partial_0(\sqrt{\gamma}DY_e) + \partial_i[\sqrt{\gamma}DY_e(\alpha v^i-\beta^i)] &=& {\alpha\sqrt{\gamma}\mathcal{R}\frac{D}{W}}. \label{eq:DY-evo}
\end{eqnarray}
\end{adjustwidth}
The last terms on the right-hand side of Equations~\eqref{eq:tau-evo}--\eqref{eq:DY-evo} are due to the NLS.

\subsection{Emissivities and Production Rates}

The classification of radiative regimes within an NS suggests a natural division between free and diffusive processes. In our scheme, the free emission rates account for the most potent reactions, including the following:

\begin{enumerate}[leftmargin=21pt,labelsep=1pt]

\item[(i)] Direct 
 Urca process, comprised of positron capture by neutrons
\begin{eqnarray}
e^+ + n \rightarrow p + \bar\nu_e,\label{eq:pc}
\end{eqnarray}
and electrons capture by protons
\begin{equation}
    e^- + p \rightarrow n + \nu_e.
\end{equation}

\item[(ii)] Electron--positron pair annihilation
\begin{equation}
e^- + e^+ \rightarrow \nu_{e} + \bar\nu_{e}\ , \qquad 
e^- + e^+ \rightarrow \nu_{\mu} + \bar\nu_{\mu}\ , 
\qquad
e^- + e^+ \rightarrow \nu_{\tau} + \bar\nu_{\tau}\ . 
\end{equation}

\item[(iii)] Transversal plasmon decay
\begin{equation}
\gamma \rightarrow \nu_{e} + \bar\nu_{e}\ , 
\qquad 
\gamma \rightarrow \nu_{\mu} + \bar\nu_{\mu}\ , 
\qquad 
\gamma \rightarrow \nu_{\tau} + \bar\nu_{\tau}. 
\label{eq:pldc}
\end{equation}
\end{enumerate}

The expressions employed to estimate the emissivities and production rates of the above processes can be found in \cite{Ruffert:1995fs}. The free emissivity rate $Q^F(I)$ and the free production rate $R^F(I)$ (with $I = \nu_e, \bar\nu_e, \nu_x$) are the sum of emission rates over the reactions \mbox{$\rm r$, that is,}
\begin{eqnarray}
Q^F(I) = \sum_{\rm r} Q_r(I),\hspace{1cm} R^F(I) = \sum_{\rm r} R_r(I). \label{eq:rfree}
\end{eqnarray}

The diffusive processes are:

\begin{enumerate}[leftmargin=21pt,labelsep=1pt]

\item[(i)] Neutrino-elastic scattering on a representative heavy nucleus $X$ and atomic mass number $A$.
\begin{eqnarray}
&&\nu_e + A \rightarrow \nu_e + A,\hspace{0.5cm}\bar\nu_e + A \rightarrow \bar\nu_e + A,\hspace{0.5cm}\nu_x + A \rightarrow \nu_x + A.
\end{eqnarray}
\item[(ii)] Neutrino-elastic scattering on free nucleons\vspace{-12pt}
\begin{adjustwidth}{-\extralength}{0cm}
\begin{eqnarray}
&&\nu_e + [n,p] \rightarrow \nu_e + [n,p],\hspace{0.5cm}\bar\nu_e + [n,p] \rightarrow \bar\nu_e + [n,p],\hspace{0.5cm}\nu_x + [n,p] \rightarrow \nu_x + [n,p].
\end{eqnarray}
\end{adjustwidth}
\item[(iii)] Electron-flavor neutrino absorption on free nucleons
\begin{equation}
\nu_e + n \rightarrow p + e^-\ , \qquad \bar\nu_e + p \rightarrow n + e^+.
\end{equation}
\end{enumerate}

The $I$ 
neutrinos mean free path $\lambda_I$, which is a function of the neutrinos energy $E_I$, is defined as
\begin{eqnarray}
    \lambda_I^{-1} &\equiv& n_p[\sigma_{I,s}(p) + \sigma_{I,a}(p)] 
    + n_n[\sigma_{I,s}(n) + \sigma_{I,a}(n)] + n_h \sigma_{I,s}(X),
\end{eqnarray}
where $n_p,~n_n,$ and $n_h$ are the protons, neutrons, and heavy nuclei number densities, respectively. The neutrino energy dependence is introduced by the scattering (subscript $s$) and absorption (subscript $a$) cross-sections found in \cite{Ruffert:1995fs}. In order to classify how opaque a medium is with respect to the $I$ neutrino, the optical depth is defined as
\begin{equation}\label{eq:tau-def}
    \tau_I(E_I) \equiv \int_{s_1}^{s_2} \frac{ds}{\lambda_I(E_I)},
\end{equation}
where the line integral above is evaluated along the invariant line element of Equation~\eqref{eq:3+1-line} with $dt = 0$ parametrized by $s$ between $s_1$ and $s_2$. Since all cross-sections used in this work depend on $E_I^2$, it is useful to factor them out in the form
$\zeta_I = (E_I^2\lambda_I)^{-1} $ and to define the energy-independent optical depth as
\begin{equation}
    \chi_I \equiv \int_{s_1}^{s_2} \zeta_I ds,\label{eq:chi}
\end{equation}
where a discussion of the method adopted to estimate $\chi_I$ is presented in Section~\ref{sec:NLS-estimates}. Finally, in terms of $\chi_I$, Equation~\eqref{eq:tau-def} reads
\begin{equation}
    \tau_I = E_I^2 \chi_I.
\end{equation}
\indent If $E_I^2$ is taken to be the ultra-relativistic Fermi--Dirac ensemble average, the expression \mbox{above becomes}
\begin{equation}
    \tau_I = \chi_I \frac{F_4(\eta_I)}{F_2(\eta_I)} T^2,
\end{equation}
for the degeneracy parameters $\eta_I = \mu_I/T$. In our implementation, the incomplete Fermi--Dirac integrals
\begin{equation}
    F_k(\eta) = \int_0^{\infty} \frac{x^k dx}{\exp(x-\eta)+1},
\end{equation}
are computed by the analytic fittings of \cite{Takahashi}. The \textit{neutrinosphere} of the $I$th neutrino is defined as the surface at which $\tau_I = 2/3$ and serves the purpose of dividing the optically thick region ($\tau_I > 2/3$), where diffusive processes dominate, from the optically thin region ($\tau_I < 2/3$), where free emission processes are more important.

Although emissivities and production rates may be estimated for diffusive and freely streaming regimes, the optical properties of NS matter with respect to neutrinos may lie in an intermediate regime.
Therefore, to capture this feature, we employ effective emissivities $Q_{\rm eff}(I)$ and effective production rates $R_{\rm eff}(I)$ at each point defined by the interpolation~\cite{Rosswog:2003rv, Galeazzi:2013mia, Foucart:2014nda}
\begin{eqnarray}
&&Q_{\rm eff}(I) \equiv Q(I) = \frac{Q^F(I)Q^D(I)}{Q^F(I)+Q^D(I)},\hspace{0.5cm} R_{\rm eff}(I) \equiv R(I) = \frac{R^F(I)R^D(I)}{R^F(I)+R^D(I)}\label{eq:Reff},
\end{eqnarray}
where the diffusive emissivity $Q^D(I)$ and the diffusive production rate $R^D(I)$ are given by~\cite{Rosswog:2003rv}

\begin{eqnarray}
&&Q^D(I) = \frac{4\pi g_I}{(hc)^3}\frac{\zeta_I}{3\chi_I^2}T^2F_1(\eta_I), \hspace{1cm} R^D(I) = \frac{4\pi g_I}{(hc)^3}\frac{\zeta_I}{3\chi_I^2}TF_0(\eta_I),\label{eq:diff-R}
\end{eqnarray}
with $g_{\nu_e} = g_{\bar\nu_e} = 1$, $g_{\nu_x} = 4$, and $h$ being the Planck constant. $Q^F(I)$ and $R^F(I)$ are given by  Equation~(\ref{eq:rfree}). Then, Equation~\eqref{eq:Reff} is used to compute the NLS contributions to the GRHD source terms from Equations~\eqref{eq:Q_tot} and \eqref{eq:R_tot}.

Finally, we estimate the source luminosity (i.e., without including redshift) for the $I$th neutrino species with the following expression~\cite{Galeazzi:2013mia}:
\begin{equation}\label{eq:luminosity}
    L_{I} = \int~d^3x~\left[\alpha \sqrt{\gamma} W n_b Q(I) \frac{(\alpha - \beta^i v_i)}{\sqrt{-g_{00}}}\right],
\end{equation}
which comes from integrating the energy per unit time measured by a coordinate observer over a refinement level.

\section{\label{sec:3} Numerical Implementation}

BAM uses a hierarchy of $L$ nested Cartesian levels labeled with \mbox{$l=0,1,\dots,L-1$}. The moving levels $l \geq l_{\rm mv}$ contain $n_{\rm mv}$ points per direction and can move to track the motion of the stars, while the static levels $l < l_{\rm mv}$ contain $n$ points per direction and are fixed. The constant distance between grid points within one level is given by \mbox{$h_l = h_0/2^l$,} where $h_0$ is the distance between grid points in level 0. The fluxes in the GRHD Equation~\eqref{eq:grhd-val-form} are estimated employing a high-resolution shock-capturing scheme based on primitives reconstruction at cell interfaces using the WENOZ scheme~\cite{BORGES20083191}, the local Lax--Friedrichs (LLF) approximate Riemann solver~\cite{Toro:2009}, and a conservative mesh refinement strategy. The time evolution is performed adopting the method of lines and a 4th-order \mbox{Runge--Kutta integrator.}

\subsection{Code Updates}

\begin{enumerate}
 
\item In 
 our previous studies using the BAM code, we used mainly one-parameter piecewise polytropes EoSs together with an ideal-gas thermal contribution. Now, we have extended this infrastructure to enable the use of three-dimensional tables. In general, these tables have a finite range of validity defined as a domain $\mathcal{D}$ with
\begin{eqnarray}
    \mathcal{D} = \{(\rho, T, Y_e) : \rho^{\min} \leq \rho \leq \rho^{\max}, T^{\min} \leq T \leq T^{\max}, Y_e^{\min} \leq Y_e \leq Y_e^{\max}\}.
\end{eqnarray}
\indent For this purpose, EoS evaluations should only be performed within this domain (i.e., additional checks have to be incorporated into BAM). 

\item Previously adopted EoSs allowed us to use a simple and fast converging root-finding procedure for the conservative-to-primitive conversion. This is not the case for a three-parameter tabulated EoS, since numerical derivatives computed by trilinear interpolations are noisy. In our case, we use the methods outlined in \cite{Galeazzi:2013mia, 2013rehy.book.....R} to ensure a robust conservative-to-primitive conversion. 

\item Once we employ three-parameter EoSs, we also have to solve Equation~\eqref{eq:DY-evo}.

\item We make use of a static and cold atmosphere to model vacuum, that is, grid points with $\rho \leq \rho_{\rm fac} \times \rho_{\rm atm}$ (here, we use $\rho_{\rm fac}=10$ and $\rho_{\rm atm} = 10\times \rho_{\min}$) are set to 
\begin{eqnarray*}
\rho = \rho_{\rm atm},\hspace{1cm} v^i = 0,\hspace{1cm} T = T^{\min} = 0.1~{\rm MeV}, \hspace{1cm} Y_e = Y_{e,\rm atm}.
\end{eqnarray*}
We use $Y_{e,\rm atm} = 0.4$ in our TOV simulations to reproduce the conditions of the testbeds reported in the literature, and $Y_{e,\rm atm} = Y^{\min} = 0.01$ in our BNS runs so that the pressure of the atmosphere $p_{\rm atm} = p (\rho_{\rm atm}, T_{\rm atm}, Y_{e,\rm atm})$ is lowest.
\end{enumerate}

\subsection{\label{sec:NLS-estimates}Free Emission Rates and Optical Depth Estimates}

Using the $\beta$-equilibrium condition, Equation~\eqref{eq:beta-eq}, and the local thermal equilibrium hypothesis allows us to compute the free emission rates $R^F(I),~Q^F(I)$ for the processes outlined in Equations~\eqref{eq:pc}--\eqref{eq:pldc} directly from the EoS. During the code initialization, we build auxiliary tables for the emission rates and $\zeta_I$; then, any required value along the simulation is computed by trilinear interpolation of the tables. Next, to calculate the effective emission rates, diffusive emission rates $R^D(I),~Q^D(I)$ must also be computed, which requires determining the energy-independent optical depth $\chi_I$ of Equation~\eqref{eq:diff-R}. Due to the lack of knowledge about the trajectory of the neutrinos within a material medium, we resort to Fermat's principle in order to choose the energy-independent optical depth, Equation~\eqref{eq:chi}, at each grid point $k$ as the minimum $\chi_I$ among the six first neighbors along the $d (= x,~y,~z)$ coordinate directions, that is,
\begin{eqnarray}
&&\chi_{I,k} = \min{\left[ \bar{\zeta}^{\pm}_{I,d}\sqrt{\bar{\gamma}^{\pm}_{dd}} \Delta x_d\right]}, \label{eq:opt-est}
\end{eqnarray}
where $\bar{\zeta}^{\pm}_{I,d}$~($\bar{\gamma}^{\pm}_{dd}$) is the average $\zeta_I$~($\gamma_{dd}$) between the point $k$ and the neighbors in the \mbox{$\pm d$ directions}, while $\Delta x_d$ is the grid spacing in the $d$ direction.

It is worth pointing out that although more elaborate approaches for the energy-independent optical depth estimation are possible (e.g., ray-by-ray integrating up to the boundaries of the computational domain~\cite{Deaton:2013sla}, using an auxiliary grid adapted to the symmetry of the system~\cite{Galeazzi:2013mia} or iteratively over the entire grid~\cite{Neilsen:2014hha, Foucart:2014nda}), such prescriptions tend to further increase computational costs and often violate special relativity.

\section{\label{sec:4} Neutrino-Induced Collapse of Single TOV Stars}

\begin{figure}[H]

\begin{minipage}{0.49\textwidth}
  
  \includegraphics[width=\linewidth]{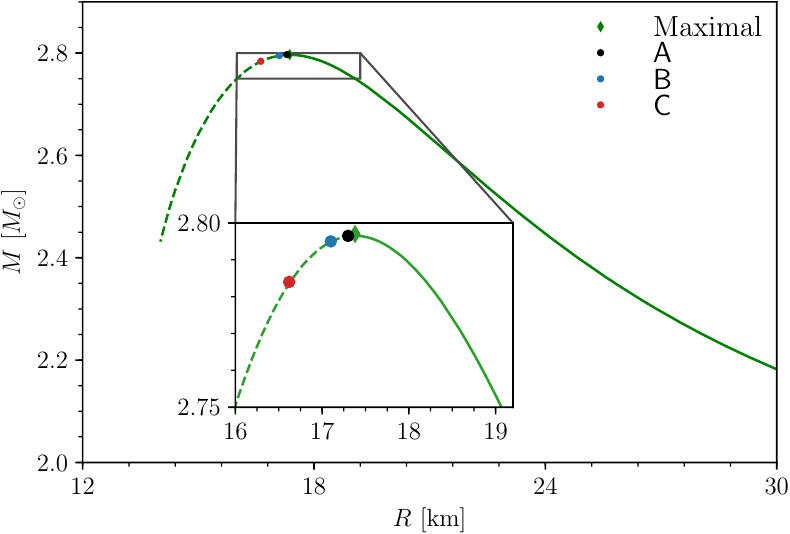}
\end{minipage}%
\begin{minipage}{0.49\textwidth}
  
  \includegraphics[width=\linewidth]{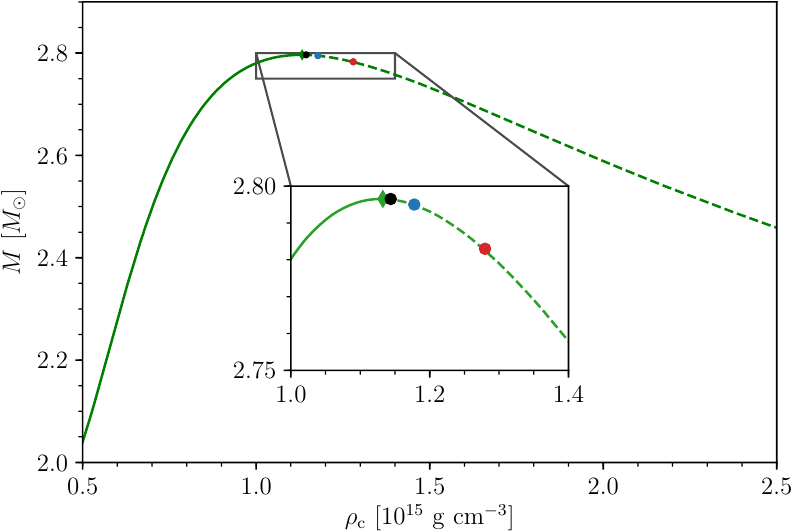}
\end{minipage}
\caption{TOV solutions for the SHT-NL3 EoS in neutrino-less $\beta$-equilibrium and constant \protect \linebreak $T = 30~{\rm MeV}$. The marked points refer to the configurations of Table~\ref{tab:1}: maximal (green diamond), A (black circle), B (blue circle), and C (red circle). {\textbf{Left} panel}: mass-radius curve. {\textbf{Right} panel}:~mass-central rest-mass density curve. The solid line represents the stable branch, while the dashed line represents the unstable branch.}
\label{fig:1}
\end{figure}

Our first aim is to test our implementations by reproducing the neutrino-induced gravitational collapse reported in \cite{Galeazzi:2013mia}. We employ the SHT-NL3 EoS~\cite{Shen:2011kr} initially in neutrino-less $\beta$-equilibrium at constant $T = 30~{\rm MeV}$. 
Integrating the TOV equations for various central rest-mass densities results in the mass-radius and mass-central rest-mass density curves depicted in Figure~\ref{fig:1}. For our simulations, we consider three radially unstable (according to the \textit{turning 
~point~criterion}~\cite{Friedman:1988er}) configurations identified by A, B, and C, with increasing rest-mass density from A to C. We evolve them with and without the NLS in a static three-level grid where the finest level encompasses the entire star. A summary of the setups is found in Table~\ref{tab:1}.

There are two main differences between our implementation and that of \cite{Galeazzi:2013mia}. First, we employ the fifth-order WENOZ reconstruction~\cite{BORGES20083191} {with LLF Riemann solver} instead of the third-order piecewise parabolic method~\cite{Colella:1982ee} {with HLLE Riemann solver.} {For completeness, we remark that the LLF scheme is a particular case of the HLLE scheme.} {Thus, although we employ a more accurate reconstruction scheme, the different choice of Riemann solver might not always ensure a smaller numerical viscosity.} Second, we estimate opacities (and hence diffusive rates) by integrating the mean free paths along the $x,~y,~z$ coordinate directions up to neighboring points.
\begin{table}[H]

\caption{Properties of the TOV stars. From left to right, the columns read: model name, central rest-mass density, gravitational mass, baryonic mass, emitted neutrino energy (up to the collapse for those evolved with NLS), grid resolution, and distance between grid points on the finest level. The central densities for this test were chosen to meet the same initial conditions of \cite{Galeazzi:2013mia}. The top row, `Maximal', refers to the model at the onset of instability.}
\newcolumntype{C}{>{\centering\arraybackslash}X}
\begin{tabularx}{\textwidth}{lm{2.5cm}<{\centering}CCm{2cm}<{\centering}m{0.8cm}<{\centering}m{1cm}<{\centering}}
\toprule
\textbf{Model} & \boldmath{$\rho_c~(10^{15}~{\rm g/cm^3})$} & \boldmath{$M~(M_\odot)$} & \boldmath{$M_b~(M_\odot)$} & \boldmath{$E~(10^{51}~{\rm erg})$} & \boldmath{$n$} & \boldmath{$h_2~({\rm m})$}\\
\midrule
Maximal & 1.068 & 2.797 & 3.506 & - & - & -\\
A & 1.079 & 2.797 & 3.310 & 2.433 & 256 & 111\\
B & 1.111 & 2.796 & 3.309 & 2.191 & 256 & 111\\
C & 1.218 & 2.784 & 3.293 & 2.075 & 256 & 111\\
\bottomrule
\end{tabularx}
\label{tab:1}

\end{table}

\begin{figure}[H]

\begin{minipage}{0.9\textwidth}
  
  \includegraphics[width=\linewidth]{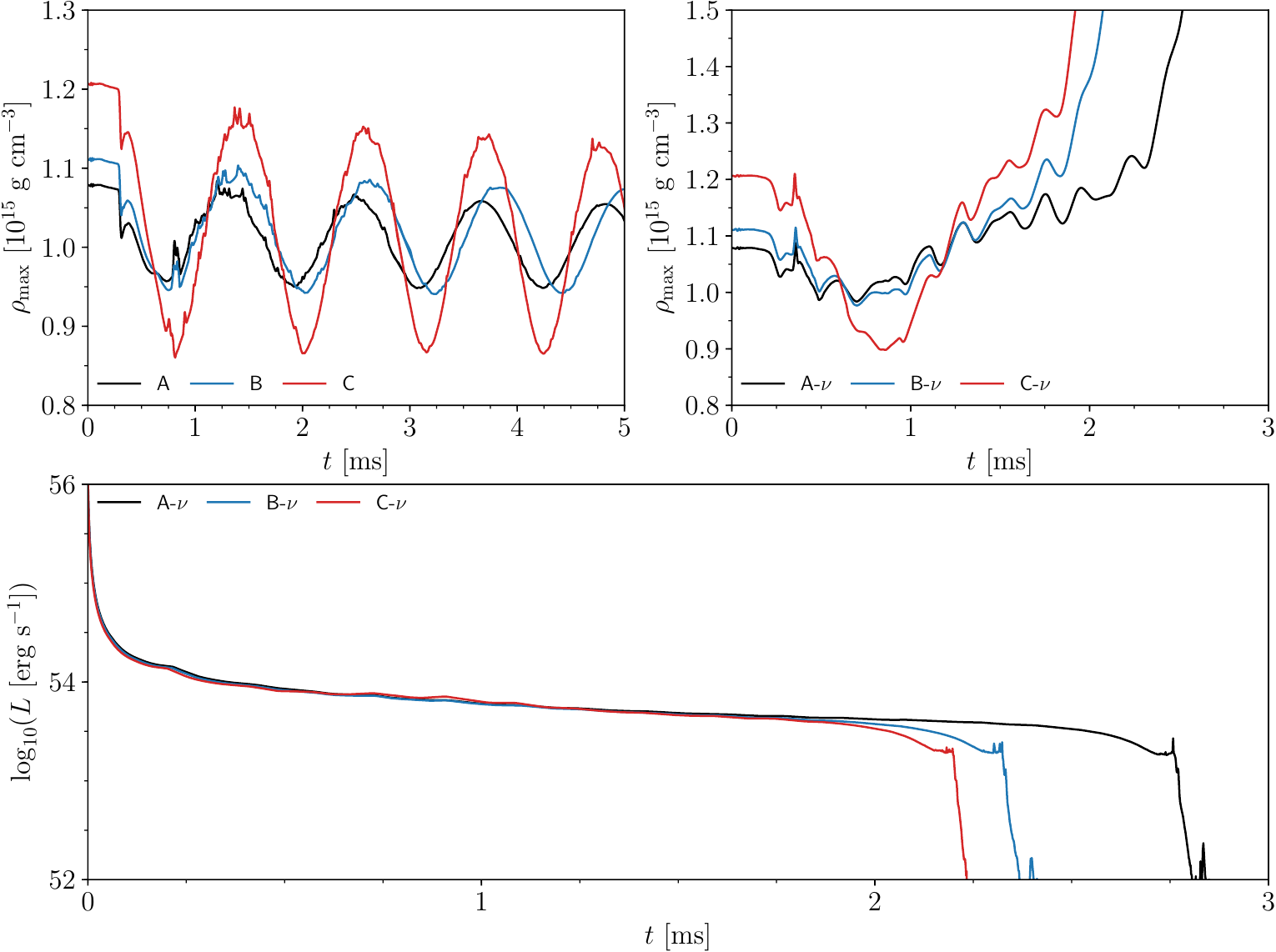}
\end{minipage}%
\caption{Time evolution of quantities of interest. {\textbf{Upper} panels}: central rest-mass density for simulations A, B, and C. {\textbf{Left}~panel}: without NLS, where the density stably evolves around an equilibrium state for each run. {\textbf{Right}~panel}: with NLS, where the wobbly evolution results from the coupling between the matter motion and the neutrinos emission. A final density growth marks the formation of an apparent event horizon. {\textbf{Lower} panel}: total luminosity evolution for simulations A-$\nu$, B-$\nu$, and C-$\nu$. We notice a rapid burst at the beginning of the simulations. Likewise, the oscillating pattern of the luminosity is a consequence of the coupling between neutrinos and matter. After the formation of the apparent event horizon, the luminosity abruptly decreases.}
\label{fig:2}
\end{figure}

In Figure~\ref{fig:2}, we present the central rest-mass density evolution for simulations without NLS (A, B, C) on the left panel and with NLS (A-$\nu$, B-$\nu$, C-$\nu$) on the right panel. During the simulations, the stars without NLS evolve stably, while NLS simulations show the characteristic density growth and gravitational collapse. 
The gravitational collapse is caused by the cooling and deleptonization that occurs more intensely in medium-low density regions of the star and leads to the decrease of the pressure exerted by those fluid elements. Unable to resist the gravitational attraction, the outer envelopes are pulled towards the dense core, decreasing the star radius and increasing the central rest-mass density. In the cases shown, the additional pressure due to the denser configuration was not enough to prevent the collapse.
In the cases with NLS, the oscillatory evolution of the rest-mass density follows from the coupling between the fluid motion and the emission of neutrinos, which are responsible for carrying away energy-momentum from the matter. Overall, the collapse takes place sooner for higher central rest-mass density since the NS is more unstable to radial oscillations, which also explains the ordering of the observed collapse time in Figure~\ref{fig:2}.

In the lower panel of Figure~\ref{fig:2}, we present the total neutrino luminosity during the simulations with NLS, where an initial burst of neutrinos is apparent due to the high initial temperature and the abundance of nucleons and electrons powering very energetic neutrino-driven reactions in semi-transparent regions of the star. The luminosity fades over time as a consequence of the rapid cooling of medium-low density material until it dips when an apparent event horizon is formed. 

\begin{figure}[H]

\begin{minipage}{0.5\textwidth}
  
  \includegraphics[width=0.9\linewidth]{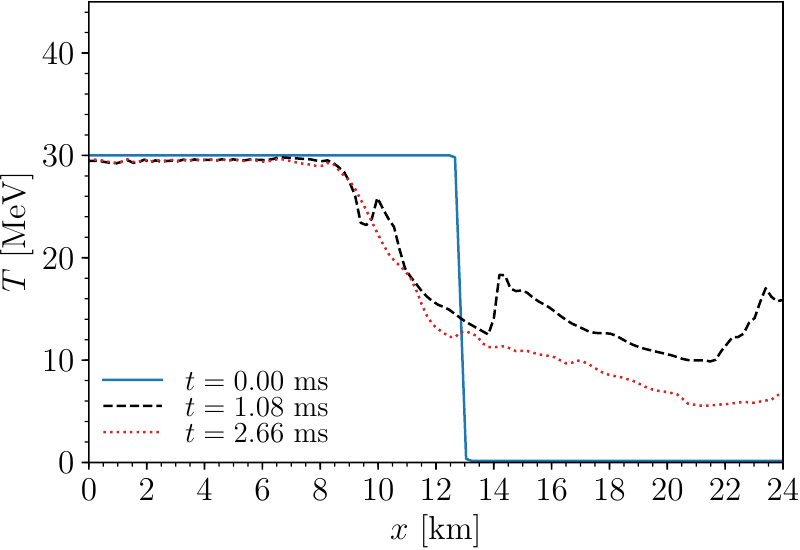}
\end{minipage}%
\begin{minipage}{0.5\textwidth}
  
  \includegraphics[width=0.9\linewidth]{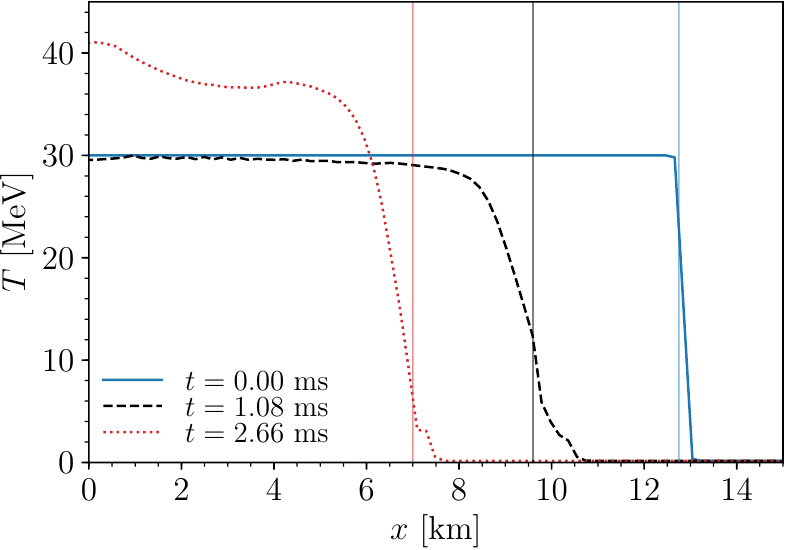}
\end{minipage}
\caption{Evolution of the temperature profile inside the star along the coordinate $x$ direction at the innermost level $l = 2$. Run A 
(\textbf{left} panel): we present temperatures up to the level $l = 2$ boundary to show that in this case, {within our simulations timespan,} the NS {has increased radius} with respect to $t = 0$. We note that at $t = 1.08~{\rm ms}$ and {$t=2.66~{\rm ms}$, the NS is still expanding and ejecting material}. Run A-$\nu$ (\textbf{right} panel): the vertical lines mark the position of the electron neutrino neutrinosphere. In general, outside the neutrinosphere, the material has lower temperatures and the cooling becomes less effective towards the core due to the dominance of diffusive processes. By the end of the first expansion cycle ($\sim$ $t=1.08~{\rm ms}$), the internal temperatures are nearly unchanged, whereas a large portion outside the neutrinosphere is cooler. Likewise, on the verge of the gravitational collapse ($t=2.66~{\rm ms}$), the layers outside of the neutrinosphere are cold, while the interior is hotter due \mbox{to compression.}}
\label{fig:4}
\end{figure}

The temperature profile evolution for runs A and A-$\nu$ are presented in Figure~\ref{fig:4} at $t = (0.00, 1.08, {2.66})~{\rm ms}$, which corresponds, respectively, to the initial configuration, the end of the first expansion cycle, and the onset of the apparent horizon detection. We observe that the $\nu_e$-neutrinosphere recedes towards the core. The outside (optically thin) regions are found effectively cooled, while inside the neutrinosphere, the dominance of diffusive processes prevents the temperature loss. At the onset of gravitational collapse {($t=2.66~{\rm ms}$)}, the internal layers are heated by compression.

\begin{figure}[H]
  
  \includegraphics[width=0.32\linewidth]{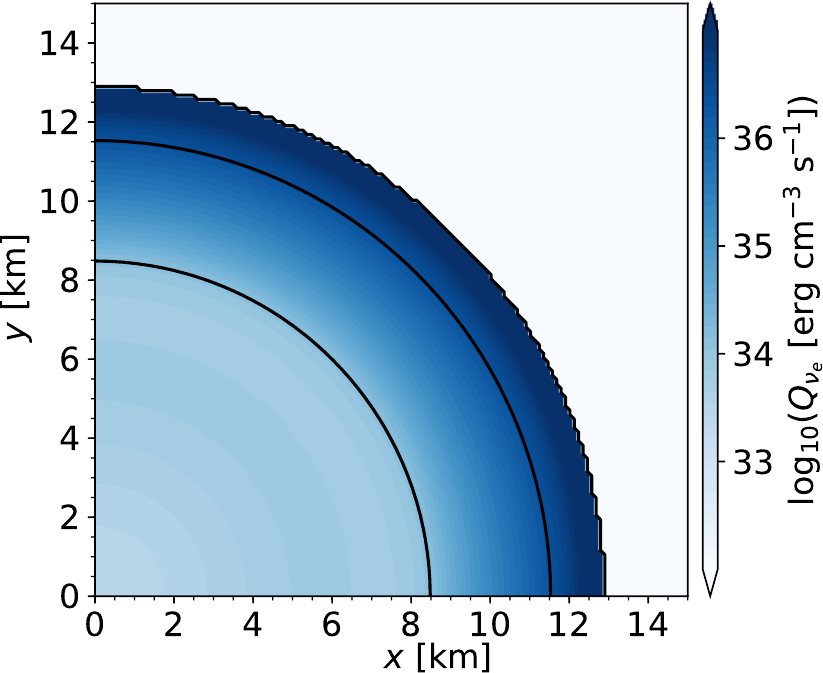}
\includegraphics[width=0.32\linewidth]{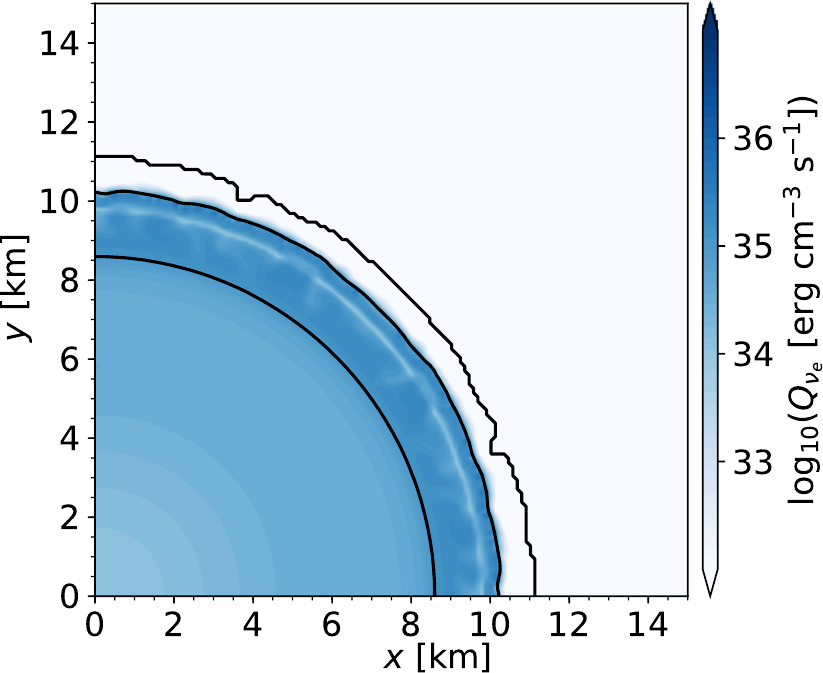}
  \includegraphics[width=0.32\linewidth]{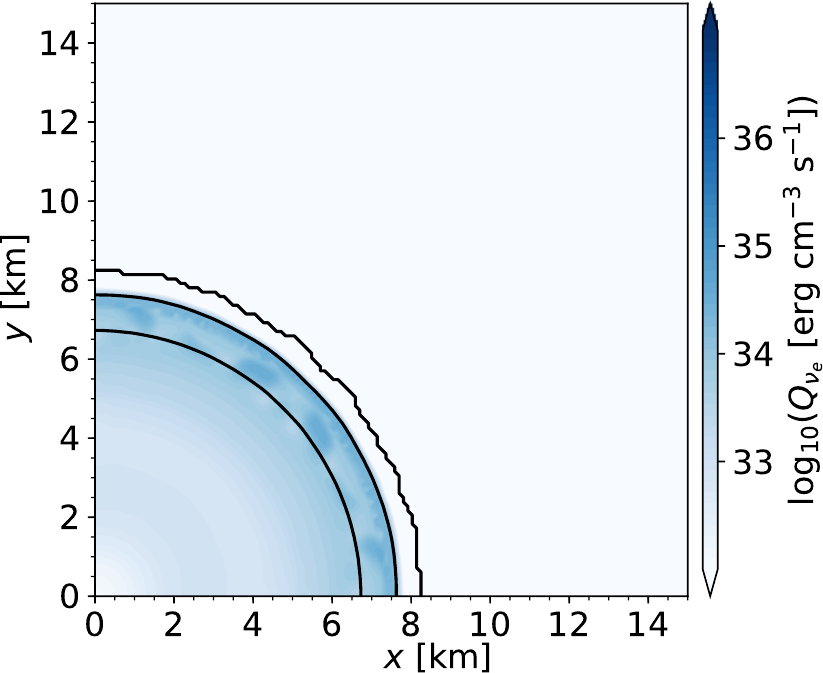}
  
  \includegraphics[width=0.32\linewidth]{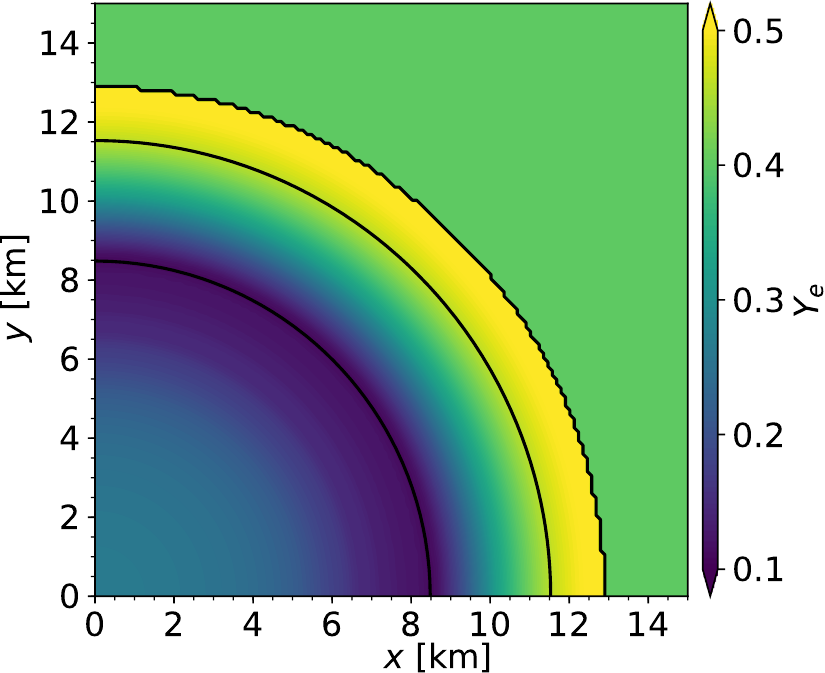}
\includegraphics[width=0.32\linewidth]{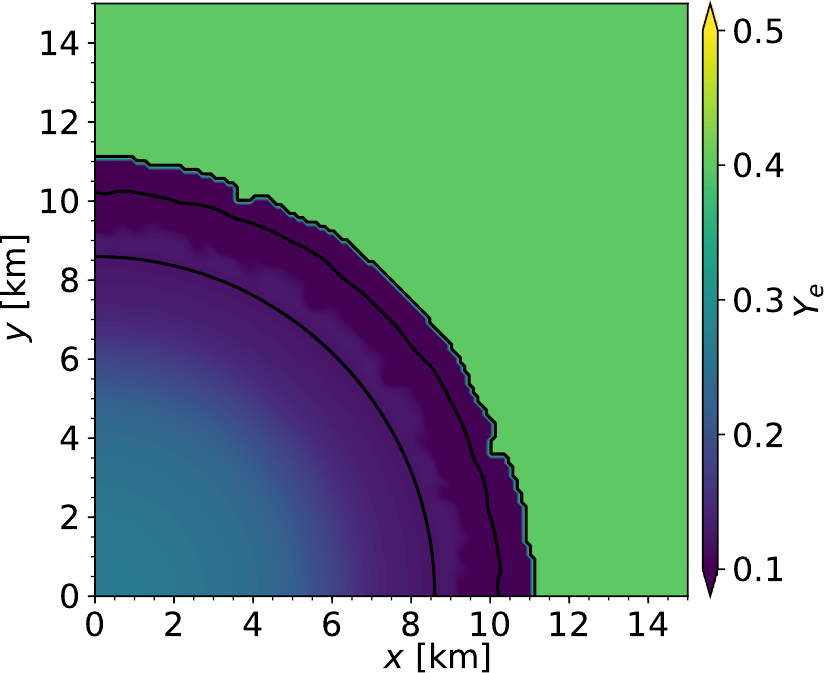}
 \includegraphics[width=0.32\linewidth]{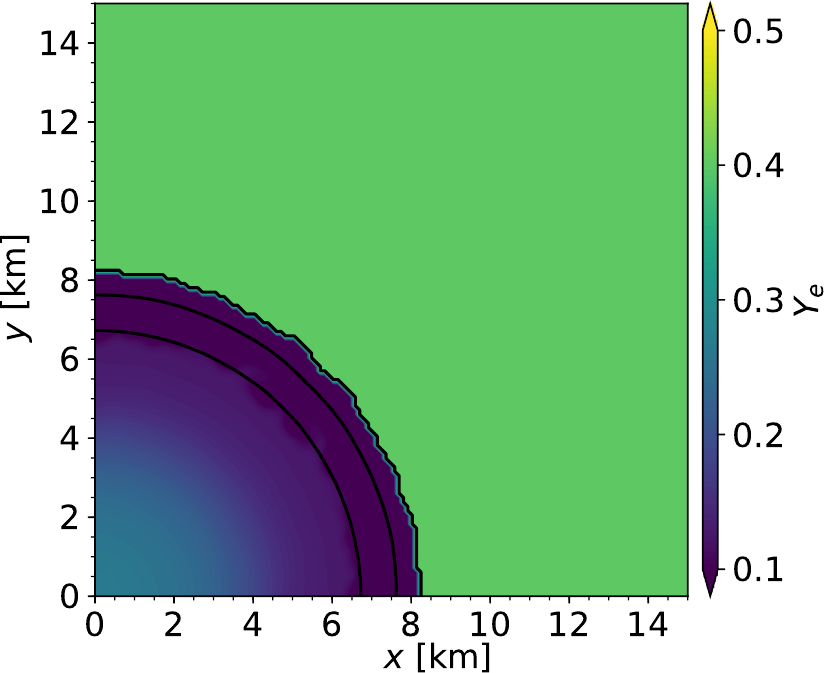}

\caption{Snapshots of the A-$\nu$ simulation at $t = 0.00,~1.08,~2.66~{\rm ms}$ from left to right.  {\textbf{Upper}~panels}: logarithm of the electron neutrinos emissivity.  {\textbf{Lower}~panels}: electron fraction. The solid black lines are contours of constant rest-mass density with $\log_{10}(\rho~[{\rm g~cm^{-3}}]) = 7, 12, 14$. The outermost line marks the interface between the star and the atmosphere.}
\label{fig:5}
\end{figure}

In Figure~\ref{fig:5}, we present snapshots of the electron neutrinos emissivity and the electron fraction for the longer-lived run A-$\nu$. We see that in the low-density envelope (with rest-mass densities between the atmosphere value $\rho_{\rm atm} = 10^7~{\rm g~cm^{-3}}$ and $\rho = 10^{12}~{\rm g~cm^{-3}}$), the emissivity decreases {by} more than {two} orders of magnitude, and the matter strongly deleptonizes (from $t=0$~ms (left panel) to $t=1.08$~ms (central panel)). This occurs within the first expansion cycle of the star and explains the early burst in the bottom panel of Figure~\ref{fig:2}. In the middle panels, the formation of eddies on a circle with radius $r = \sqrt{x^2 + y^2} \sim 9~{\rm km}$ is related to convective instabilities as predicted by the Ledoux criterion~\cite{Epstein:1978ih}. On the right panels, the star is on the verge of gravitational collapse. Note that along the evolution, the emissivity and electron fraction is almost unchanged within the opaque region $\rho \geq 10^{14}~{\rm g/cm^3}$ because much less energetic diffusive processes dominate the emissions. 

\vspace{-6pt}

\section{\label{sec:5} BNS Simulations}

In order to compare the performance of our implementations in the case of BNSs with those reported in the literature, in particular \cite{Foucart:2015gaa}, we first performed short inspiral (\mbox{$\sim$2 to 4 orbits}), single-resolution simulations for equal-mass, non-spinning BNSs described by the SFHo~\cite{Steiner:2012rk} and the DD2~\cite{Hempel:2009mc} EoSs, with and without our NLS implementation. Both EoSs contain neutrons, protons, electrons, and positrons, initially at constant \mbox{$T = 0.1~{\rm MeV}$} and in $\beta$-equilibrium. Besides the goal of comparing results, DD2 is a stiff EoS, and SFHo is a soft EoS; hence, it is useful to assess our code capability to handle NSs belonging to both higher and lower compactness.

Furthermore, in order to evaluate the accuracy of our code, we performed long inspiral simulations ($\sim$14 orbits) using the BHB$\Lambda \phi$ EoS~\cite{Banik:2014qja} (which contains protons, neutrons, electrons, positrons, the $\Lambda_0$ hyperon, and the $\phi$ meson) with four different grid resolutions, with and without our NLS implementation. We choose this EoS to test our code with a microphysical description that contains a transition between pure nucleonic matter to hyperonic matter at high densities, thus representing a notably distinct scenario compared to that addressed by DD2 and SFHo. We consider the case of an equal-mass, non-spinning BNS system starting with a $\beta$-equilibrated, isentropic configuration, with entropy per baryon $s \sim 1=k_B$. All the EoS tables used in this work were obtained from the CompOSE online repository~\cite{Typel:2013rza, CompOSE}.

The grids used in all simulations have $L = 7$ refinement levels, and the number of moving levels is set to $l_{\rm mv} = 4$. Relevant information about grid and matter setups of our runs without the NLS are found in Table~\ref{tab:2}. We identify the results of the NLS runs by appending the suffix-$\nu$ to the simulation name. In Appendix~\ref{app:A}, we present a resolution study for the BHB$\Lambda \phi$ setup.

\subsection{Initial Data}
The initial data (ID) for our simulations were constructed using the pseudospectral SGRID code~\cite{Tichy:2009yr, Tichy:2012rp, Dietrich:2015pxa, Tichy:2019ouu}, which solves the 3 + 1 constraint equations in the conformal thin-sandwich approach~\cite{Wilson:1995uh,Wilson:1996ty,York:1998hy} by adopting a surface-fitting strategy. When this project started, SGRID only supported one-dimensional piecewise polytropic (pwp) EoSs. Therefore, we reduce the three-dimensional EoS to one dimension by imposing (i) neutrino-less $\beta$-equilibrium and (ii) either constant temperature or constant entropy per baryon $s$. Then, we parametrize the resulting one-dimensional table as a pwp, adopting a similar procedure as that of \cite{Read:2008iy}. In order to validate our approach, we point out that the TOV solutions obtained with the one-dimensional tables and the corresponding pwps have maximum differences in the coordinate radii of $\sim$ $0.08\%$ and in the tidal deformabilities~\cite{Hinderer:2007mb} of $\sim$ $1.2\%$ for our cases. For the longer runs, the BHB$\Lambda \phi$ EoS, an eccentricity reduction procedure, was employed~\cite{Kyutoku:2014yba}.
\begin{table}[H]
\caption{Binary neutron star simulations. From left to right, the columns read: simulation name, gravitational mass of the stars ($A,~B$) in isolation, baryonic mass of the stars, compactness of the stars, tidal deformability of stars, ADM mass and ADM angular momentum of the BNS at the beginning of the simulation, initial coordinate distance between the stars, number of points per direction on the static levels, number of points per direction on the moving levels, and grid spacing at the finest level.}
	\begin{adjustwidth}{-\extralength}{0cm}
		\newcolumntype{C}{>{\centering\arraybackslash}X}
		\begin{tabularx}{\fulllength}{lCCm{0.5cm}<{\centering}CCCm{1cm}<{\centering}m{0.4cm}<{\centering}m{0.5cm}<{\centering}m{1cm}<{\centering}}\toprule
\textbf{Model} & \boldmath{$M^{A,B}~[M_\odot]$} & \boldmath{$M_b^{A,B}~[M_\odot]$} & \boldmath{$\mathcal{C}^{A,B}$} & \boldmath{$\Lambda^{A,B} [\times 10^3]$} & \boldmath{$M_{\rm ADM}~[M_\odot]$} & \boldmath{$J_{\rm ADM}~[M_\odot^2]$} & \boldmath{$d_0~[{\rm km}]$} & \boldmath{$n$} &\boldmath{$n_{\rm mv}$} & \boldmath{$h_6~[{\rm km}]$}\\
\midrule
DD2 & $1.200$ & $1.292$ & $0.134$ & $1.616$ & $2.375$ & $5.612$ & $36.2$ & $256$ & $128$ & $0.199$ \\
SFHo & $1.200$ & $1.300$ & $0.148$ & $0.860$ & $2.376$ & $5.673$ & $38.0$ & $256$ & $128$ & $0.186$\\
BHB$\Lambda \phi - {\rm R}1$  & $1.350$ & $1.458$ & $0.144$ & $0.944$ & $2.679$ & $8.021$ &$58.8$ & $128$ & $64$ & $0.417$ \\
BHB$\Lambda \phi - {\rm R}2$  & $1.350$ & $1.458$ & $0.144$ & $0.944$ & $2.679$ & $8.021$ & $58.8$ & $192$ & $96$ & $0.278$ \\
BHB$\Lambda \phi - {\rm R}3$  & $1.350$ & $1.458$ & $0.144$ & $0.944$ & $2.679$ & $8.021$ & $58.8$ & $256$ & $128$ & $0.209$ \\
BHB$\Lambda \phi - {\rm R}4$  & $1.350$ & $1.458$ & $0.144$ & $0.944$ & $2.679$ & $8.021$ & $58.8$ & $320$ & $160$ & $0.167$ \\
			\bottomrule
		\end{tabularx}
	\end{adjustwidth}
\label{tab:2}
\end{table}

\subsection{Short Inspiral Simulations}

In order to test our code and compare results with those of \cite{Foucart:2015gaa}, we performed short inspiral simulations using the DD2, DD2-$\nu$, SFHo, and SFHo-$\nu$, focusing on the {most energetic} $(l=2,m=2)$ mode 
of the GW and features of the post-merger stage, where the larger differences between the different cases appear. {In NR, it is usual to decompose the GW signal in terms of $s = -2$ spin-weighted spherical harmonics $_{-2}Y_{lm}(\theta,\phi)$, such that the GW strain reads $h(t,r,\theta,\phi) = \sum_{l = 2}^{l_{\rm max}}\sum_{m=-l}^{l}h_{lm}(t,r)~_{-2}Y_{lm}(\theta,\phi)$, where $(r,\theta,\phi)$ are usual spherical polar coordinates defined in terms of the grid coordinates $(x,y,z)$.} 

The inspiral stage in the SFHo and SFHo-$\nu$ simulations is composed of $\approx$ $4$ orbits and took $t_{\rm mrg} \approx 10.3$~ms to merge, while the DD2 and DD2-$\nu$ simulations are composed of $\approx$ $2$ orbits and took $t_{\rm mrg} \approx 6.2$~ms to merge. We consider the merger time as the time at which the amplitude of the $(2,2)$ mode of the GW has its maximum.

\subsubsection{Post-Merger Stage}

\begin{figure}[H]

\includegraphics[width=0.5\textwidth]{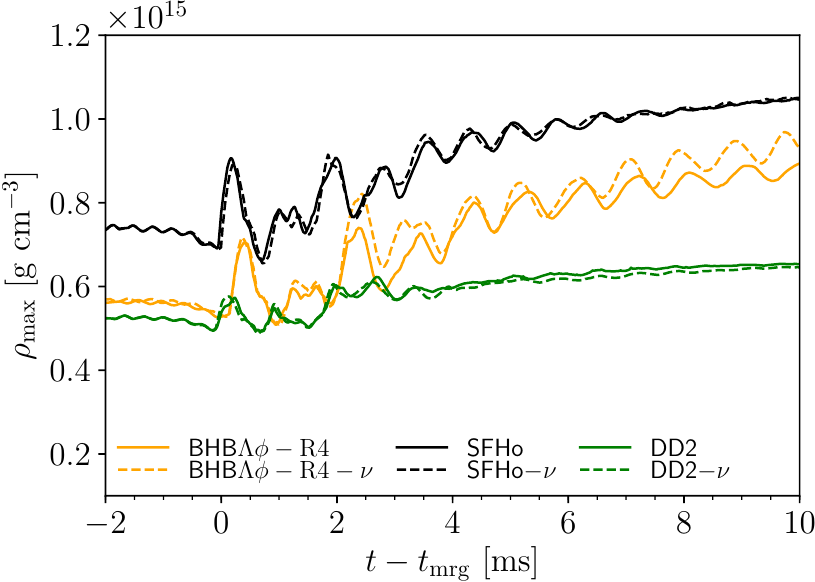}
\caption{Post-merger evolution of the maximum rest-mass density $\rho_{\max}$ reached by the different EoSs considered in this article. Solid lines refer to the runs without NLS, while dashed lines represent the NLS runs. Our results for the SFHo and DD2 cases are in agreement with those reported in~\cite{Foucart:2015gaa}. For the BHB$\Lambda \phi$ EoS, we use resolution R4 (see Table~\ref{tab:2}). }
\label{fig:6}
\end{figure}

In Figure~\ref{fig:6}, we present the maximum rest-mass density evolution during the post-merger phase of our simulations. The DD2 and SFHo cases agree with those reported in \cite{Foucart:2015gaa}, and little difference is introduced by the adoption of the NLS. The softer SFHo undergoes a merger marked by an abrupt variation of the rest-mass density around $t - t_{\rm mrg} = 0$ and a considerable increment over the whole remnant evolution. The stiffer DD2 case also exhibits these features, but in a mild way. Furthermore, in the figure, we present the BHB$\Lambda \phi$ EoS, which is an intermediate EoS located between the soft SFHo and the stiff DD2. This last mentioned EoS will be discussed in Section~\ref{sec:5.3}.

\begin{figure}[H]

\begin{minipage}{0.24\textwidth}
  
  \includegraphics[width=\linewidth]{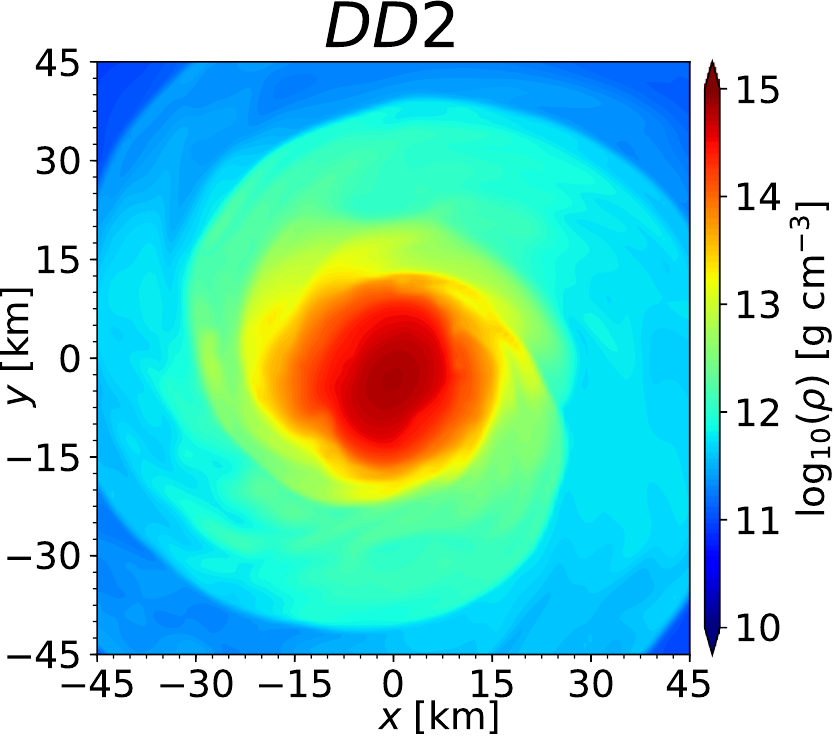}
\end{minipage}%
\begin{minipage}{0.24\textwidth}
  
  \includegraphics[width=\linewidth]{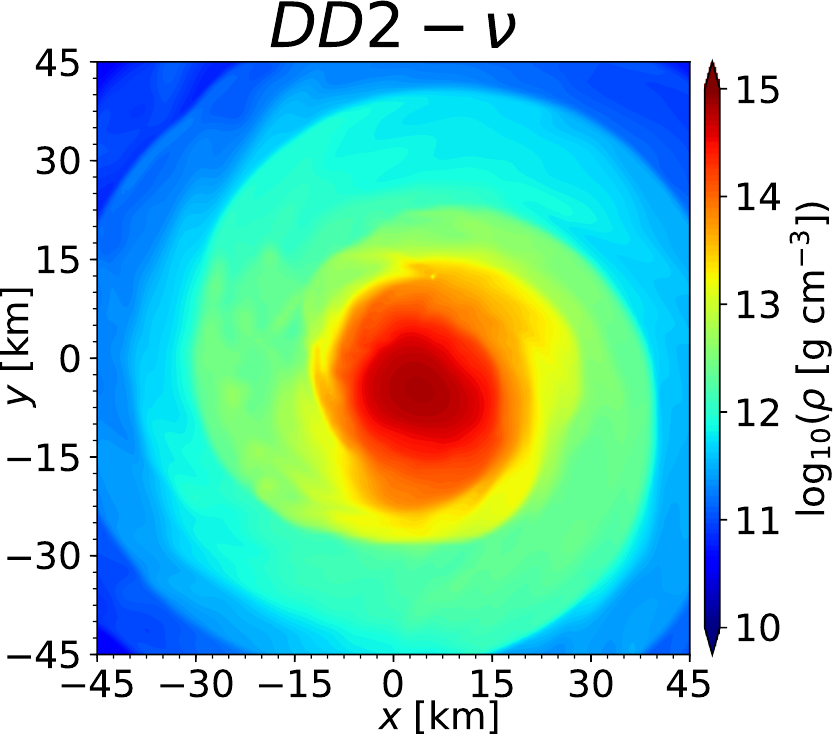}
\end{minipage}
\begin{minipage}{0.24\textwidth}
  
  \includegraphics[width=\linewidth]{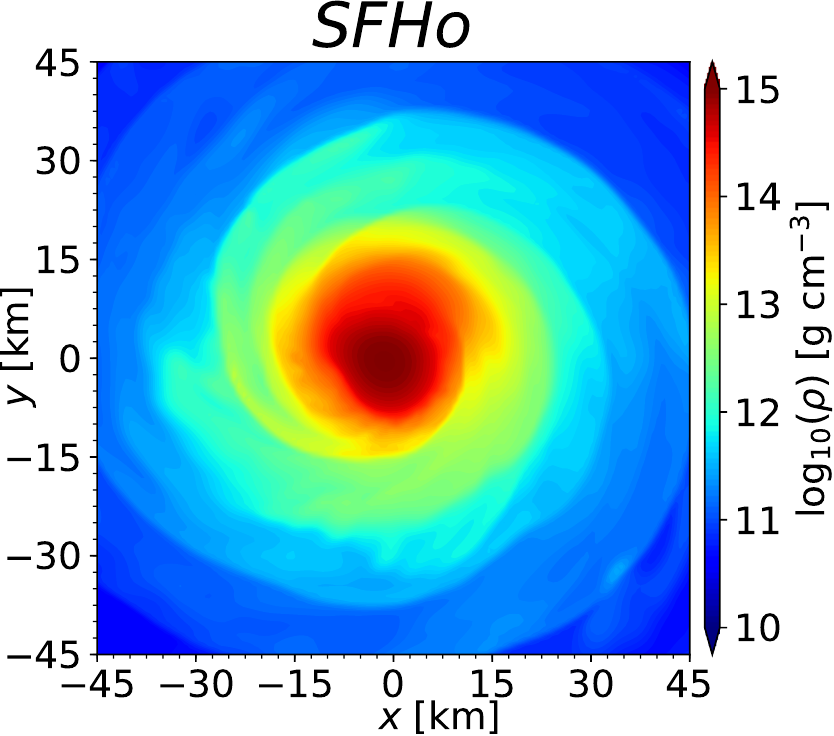}
\end{minipage}
\begin{minipage}{0.24\textwidth}
  
  \includegraphics[width=\linewidth]{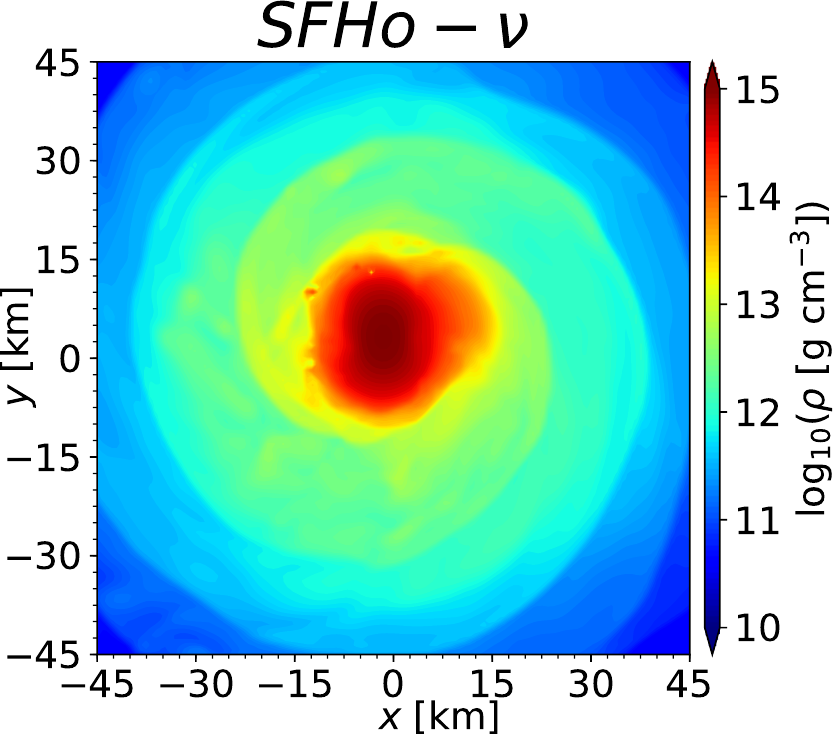}
\end{minipage}
\begin{minipage}{0.24\textwidth}
  
  \includegraphics[width=\linewidth]{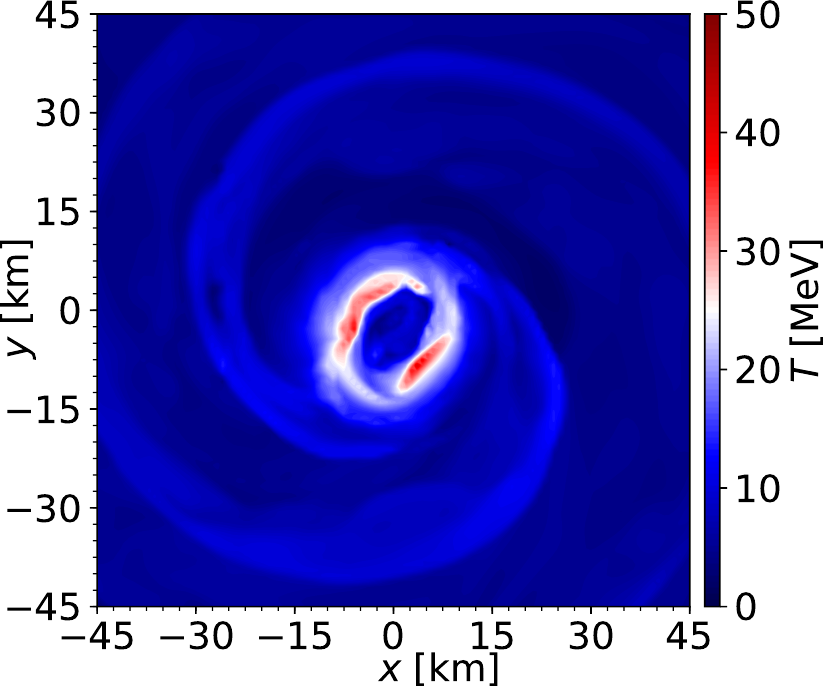}
\end{minipage}%
\begin{minipage}{0.24\textwidth}
  
  \includegraphics[width=\linewidth]{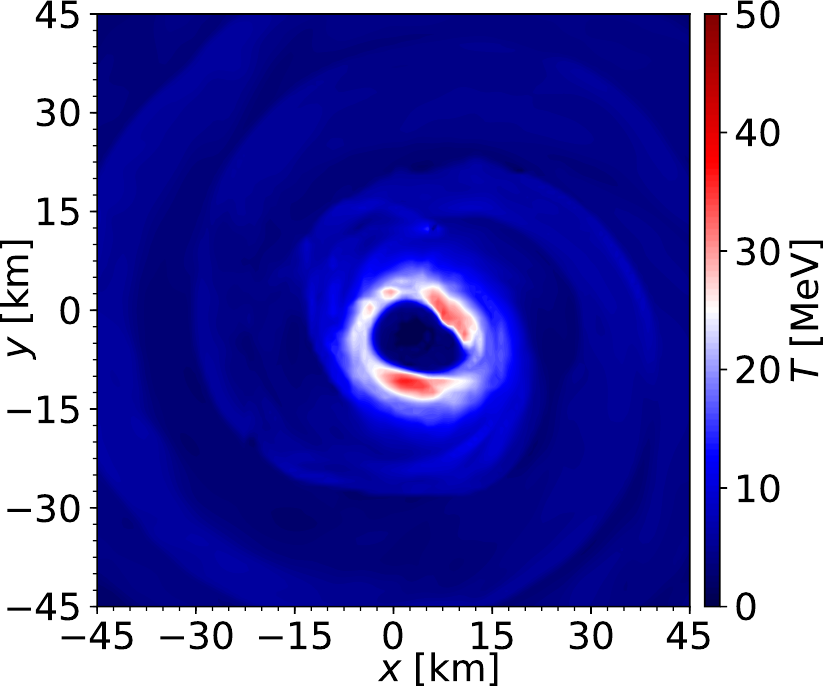}
\end{minipage}
\begin{minipage}{0.24\textwidth}
  
  \includegraphics[width=\linewidth]{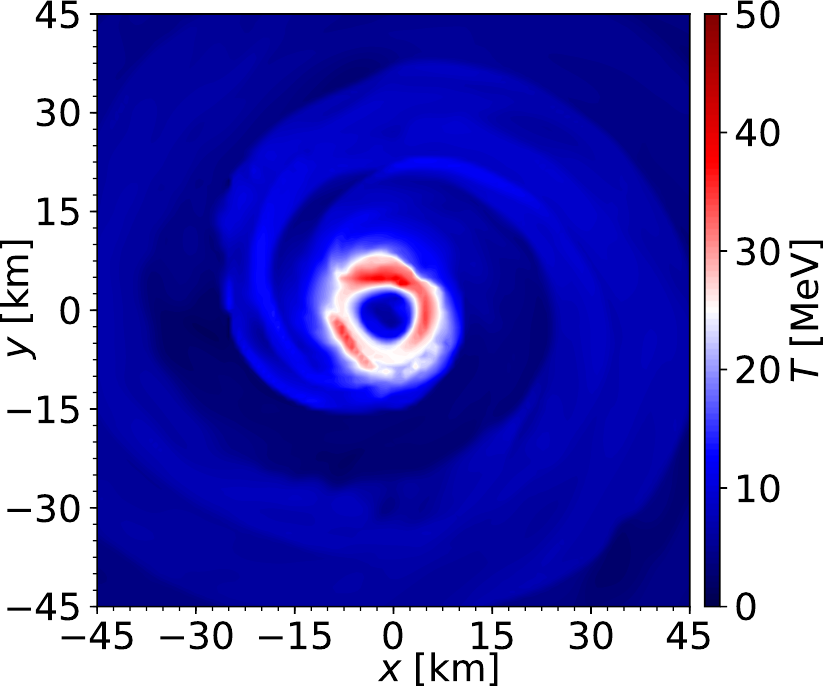}
\end{minipage}
\begin{minipage}{0.24\textwidth}
  
  \includegraphics[width=\linewidth]{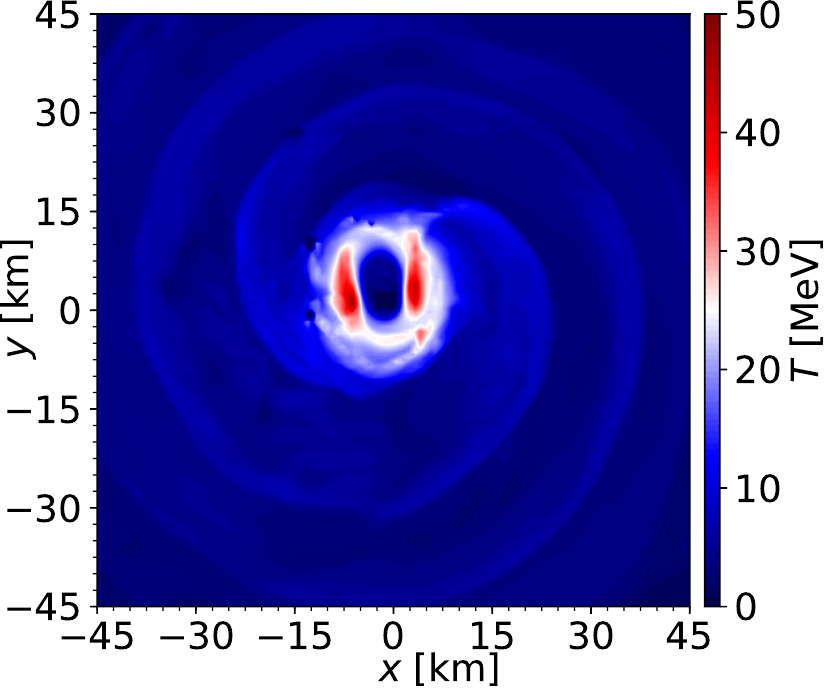}
\end{minipage}
\begin{minipage}{0.24\textwidth}
  
  \includegraphics[width=\linewidth]{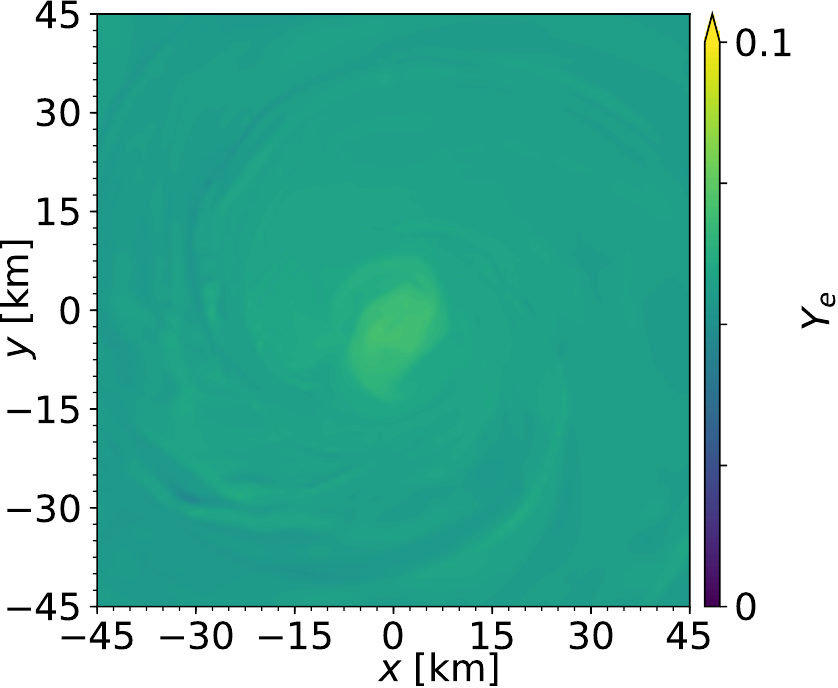}
\end{minipage}%
\begin{minipage}{0.24\textwidth}
  
  \includegraphics[width=\linewidth]{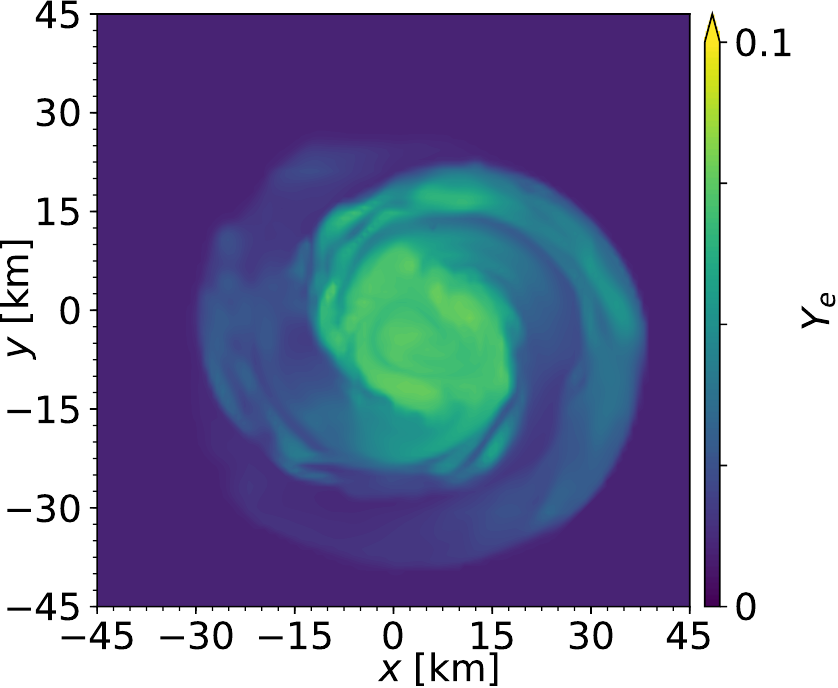}
\end{minipage}\hspace{0.2cm}
\begin{minipage}{0.24\textwidth}
  
  \includegraphics[width=\linewidth]{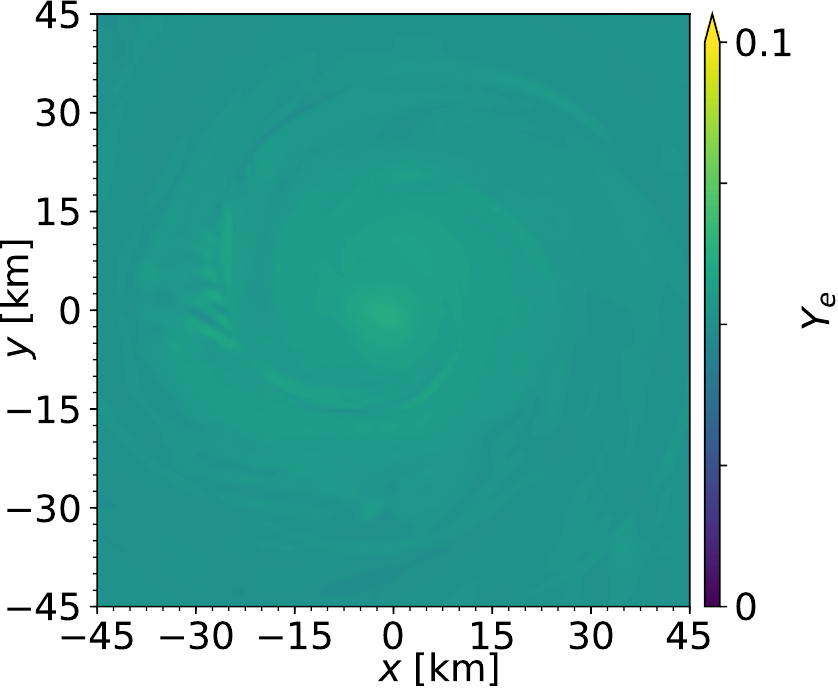}
\end{minipage}\hspace{0.2cm}
\begin{minipage}{0.24\textwidth}
  
  \includegraphics[width=\linewidth]{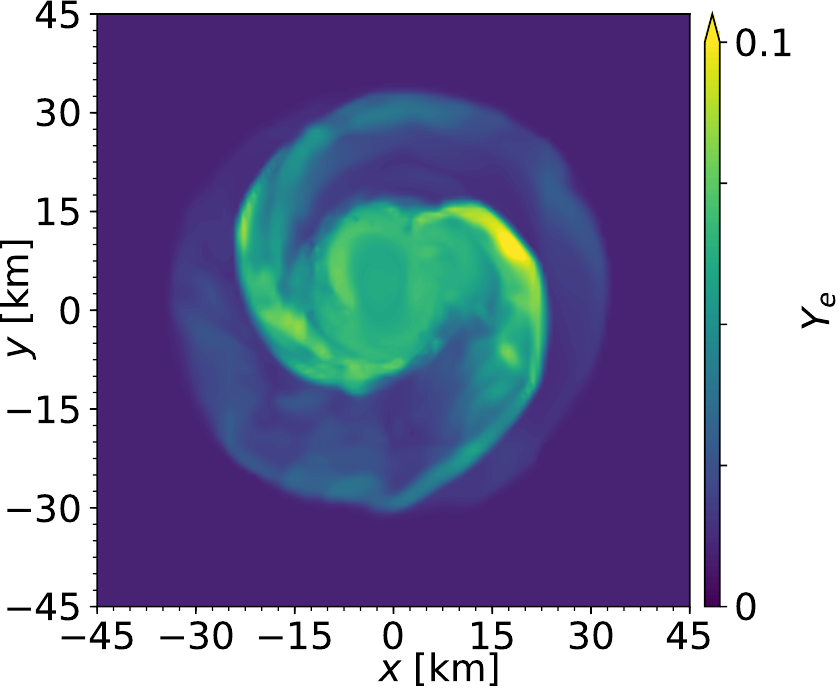}
\end{minipage}
\caption{Snapshots of the remnant $10~{\rm ms}$ after the merger in the $x$-$y$~plane. From left to right: DD2, DD2-$\nu$, SFHo, and SFHo-$\nu$. From top to bottom: rest-mass density, temperature, and electron fraction. Overall, we note the formation of bar-like structures (see the extended $\rho \geq 10^{13}~{\rm g~cm^{-3}}$ central regions and the $\rho \geq 10^{12}~{\rm g~cm^{-3}}$ arms in the top panels) surrounded by dense disks. The temperature profiles exhibit a hot interface between the bar and the disk. Finally, the effects of the NLS are perceivable on the electron fraction, where the disk becomes more neutron-rich, as opposed to the cores and spiral arms.}
\label{fig:7}
\end{figure}

In Figure~\ref{fig:7}, we show the rest-mass density, temperature, and electron fraction for the DD2 and SFHo EoSs (with and without the NLS) at $10~{\rm ms}$ after the merger. Similar to the results of \cite{Foucart:2015gaa}, we notice the formation of bar structures (the `smeared out' $\rho \geq 10^{13}~{\rm g~cm^{-3}}$ region and the extended $\rho \lesssim 10^{13}~{\rm g~cm^{-3}}$ arms on the top panels), surrounded by dense disks. The difference in compactness is also visible, since DD2 develops a less dense core than SFHo and its disk extends further. Likewise, in the middle panels, hot interfaces between the colder bar and the disk are formed, with larger maximum temperatures reached by the softer SFHo EoS. Overall, the rest-mass density and temperature are less affected by the adoption of the NLS. A noticeable difference, however, is present on the electron fraction of the remnant (bottom panels), where the outer regions of the disk become neutron-rich, while the core and arms are more leptonized. We were not able to reproduce the $Y_e \sim 0.2,~0.3$ of \cite{Foucart:2015gaa} at the outer regions of the disk, which may be attributed to a key difference in our NLS implementation (see \cite{Foucart:2014nda} for details): our optical depths are estimated from the minimum among first neighbors (as of Equation~\eqref{eq:opt-est}), instead of integrating from the current position up to a computational domain boundary along minimum optical paths. This leads to smaller optical depths in our simulations and stronger deleptonization of the disk. 
\vspace{-6pt}

\subsubsection{Spectrograms}

\begin{figure}[H]

\begin{minipage}{0.3\textwidth}
  
  \includegraphics[width=\linewidth]{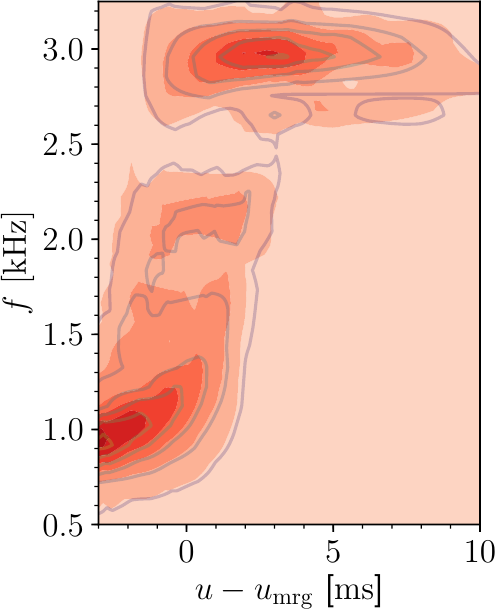}
\end{minipage} \hspace{1cm}
\begin{minipage}{0.3\textwidth}
  
  \includegraphics[width=\linewidth]{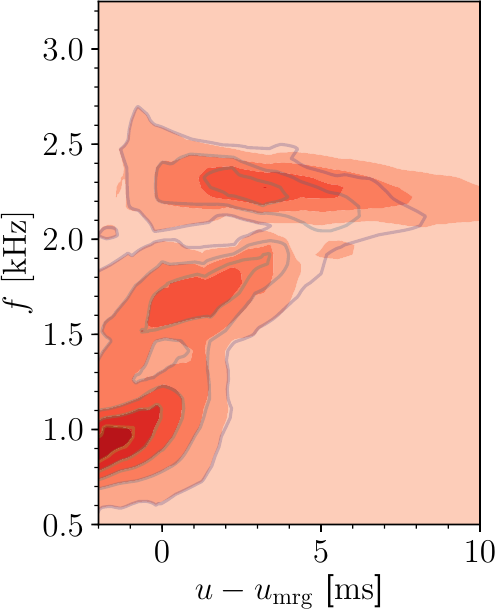}
\end{minipage}
\caption{Spectrograms of the GWs for optimally oriented binaries and extracted at $r_{\rm ext} = 600~M_\odot$. The countour lines represent the simulations without NLS. {{\textbf{Left}}~panel}: SFHo-$\nu$ simulation, where the dominant post-merger frequency is 2.95~kHz. {{\textbf{Right}}~panel}: DD2-$\nu$ simulation, where post-merger peak frequency is $\approx$2.4~kHz. Both panels share the same properties as presented {in}~\cite{Foucart:2015gaa}.}
\label{fig:8}
\end{figure}

In Figure~\ref{fig:8}, the spectrograms of the GWs of the DD2-$\nu$ and SFHo-$\nu$ runs (see \cite{Chaurasia:2018zhg} for details on the computation of the spectrograms) are presented in order to compare features of the post-merger GW signals with those reported in \cite{Foucart:2015gaa}. Our results are presented with respect to the retarded time $u$, given by
\begin{equation*}
    u = t - r_{\rm ext} - 2M \ln(r_{\rm ext}/2M - 1),
\end{equation*}
where we choose $r_{\rm ext} = 600~M_\odot$, and $M = M^A + M^B$ is the total gravitational mass of the system. As expected, we find that the NLS has little effect on the emitted GW signals, as can be seen in the similarity between the filled red contours (NLS simulations) and gray contour lines (simulations without NLS) of Figure~\ref{fig:8}.

The SFHo spectrogram reproduces the same features as the one presented in \cite{Foucart:2015gaa}. Among these features are the strongest peak during post-merger, which is clearly visible at frequency 2.95~kHz, and a frequency gap between $\approx$[2.0, 2.7]~kHz. 
In relation to the DD2 spectrogram, it also reproduces the global features, with a gap between $\approx$[2.0, 2.2]~kHz and a stronger peak at $\approx$2.4~kHz.

\subsubsection{Neutrinos Emission}

\begin{figure}[H]

\begin{minipage}{0.48\textwidth}
  
  \includegraphics[width=\linewidth]{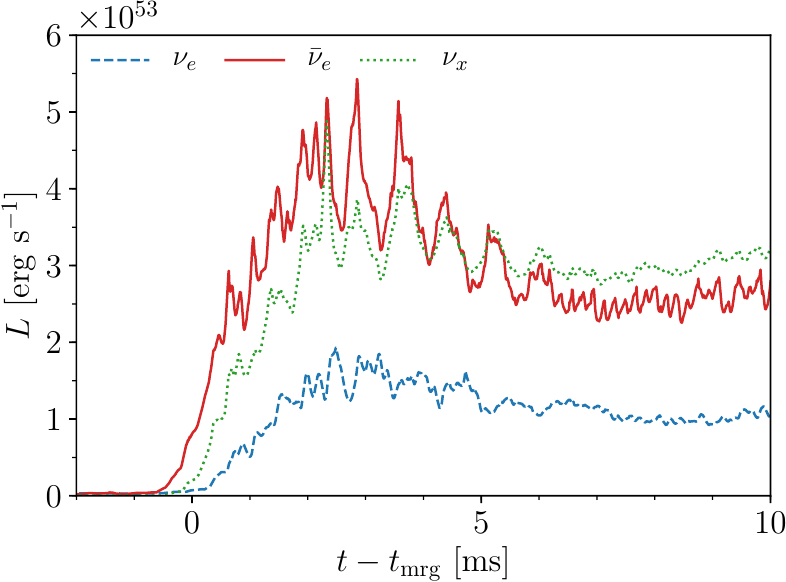}
\end{minipage}
\begin{minipage}{0.48\textwidth}
  
  \includegraphics[width=\linewidth]{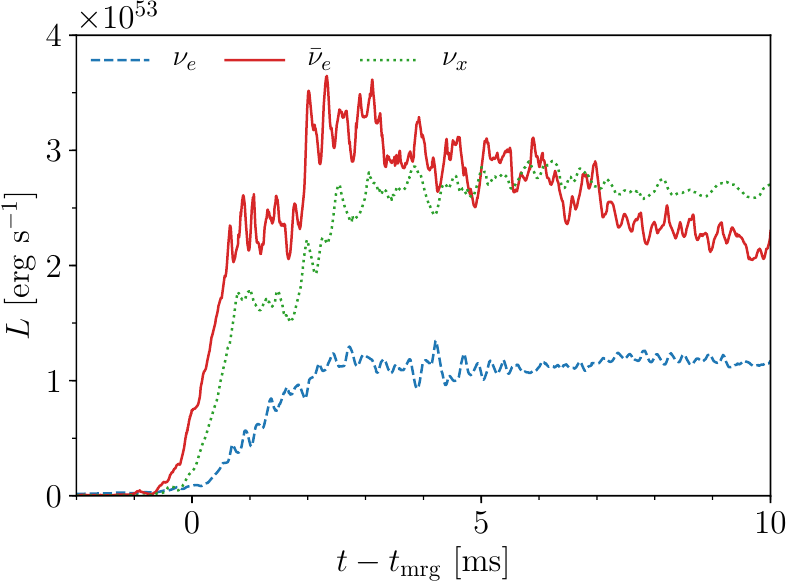}
\end{minipage}
\caption{Neutrinos source luminosity evolution for the SFHo-$\nu$ simulation ({\textbf{left} panel}) and the DD2-$\nu$ simulation ({\textbf{right} panel}). The electron antineutrinos have higher luminosity until $\sim$ $5~{\rm ms}$ after the merger.}
\label{fig:9}
\end{figure}

As depicted in Figure~\ref{fig:9}, we see that the source luminosities, Equation~\eqref{eq:luminosity}, are negligible during the inspiral stage and increase towards the merger, when the compression of matter elements leads to an increase in temperature. In both panels, we observe peaks for the three species around $2$ to $3~{\rm ms}$ after the merger, with the electron antineutrinos dominating up to $t-t_{\rm mrg} \sim 5~{\rm ms}$, and then an overall decrease towards the end of the simulation. This behavior is consistent with \cite{Foucart:2015gaa}, which then states that at $t-t_{\rm mrg} = 10~{\rm ms}$, $L_{\bar\nu_e} \sim (2 - 3)\times10^{53}~{\rm erg s^{-1}}$, with $L_{\bar\nu_e}$ dominating $L_{\nu_e}$ by a factor of 1.4--2. In our case, the electron antineutrinos luminosity lies within the same range and dominates the electron neutrinos luminosity by a factor of 1.7--2.3. Such an agreement is as good as we could expect for a leakage scheme. Concerning the total luminosity $L_{\rm tot} = L_{\nu_e} + L_{\bar\nu_e} + L_{\nu_x}$, we find systematically higher values at the peaks than those reported in \cite{Foucart:2015gaa}. At the end of our simulations, we have $L_{\rm tot} = 6.9\times10^{53}~{\rm erg~s^{-1}}$ for our SFHo run, which deviates {by} less than $1\%$ with respect to the counterpart of \cite{Foucart:2015gaa}, and $L_{\rm tot} = 5.8\times10^{53}~{\rm erg~s^{-1}}$ for the DD2 run, with a larger deviation of $\sim$ $20\%$.

\begin{figure}[H]

  \includegraphics[width=0.48\linewidth]{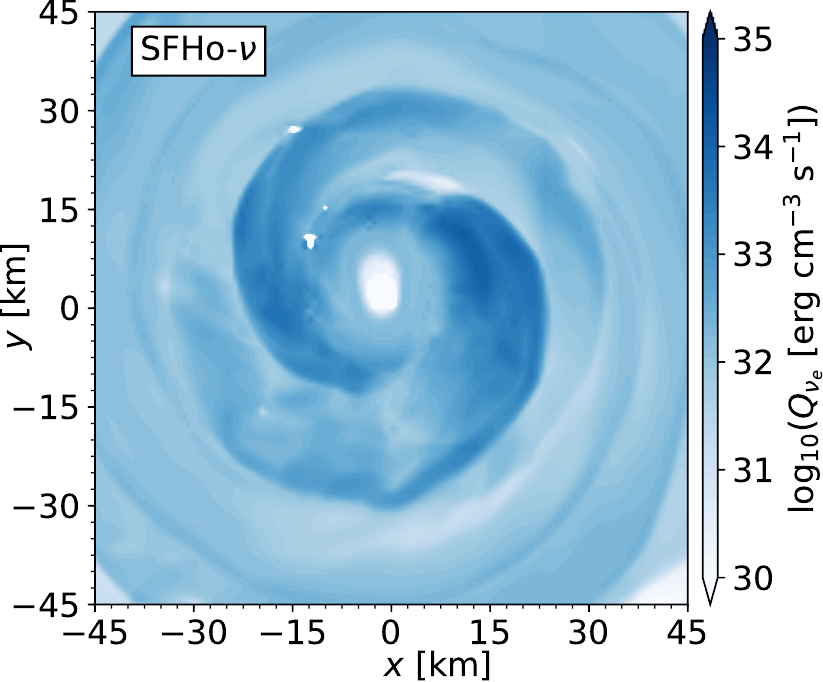}  
  \includegraphics[width=0.48\linewidth]{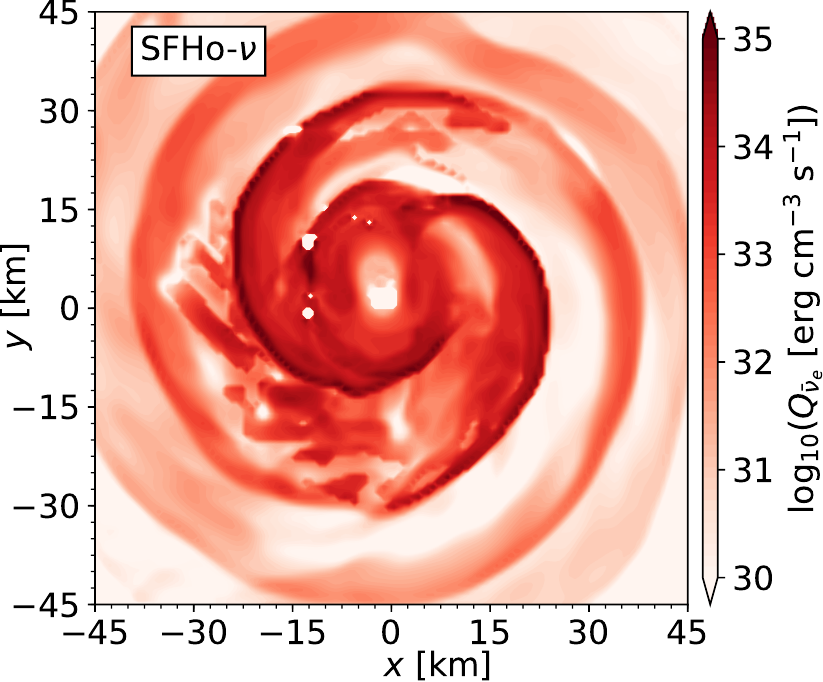}
  
  \includegraphics[width=0.48\linewidth]{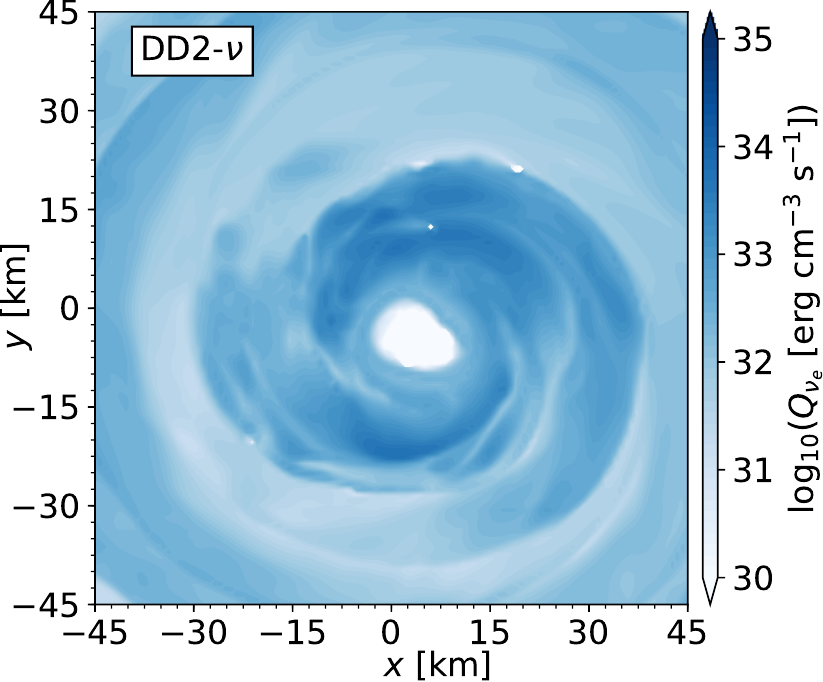}
  \includegraphics[width=0.48\linewidth]{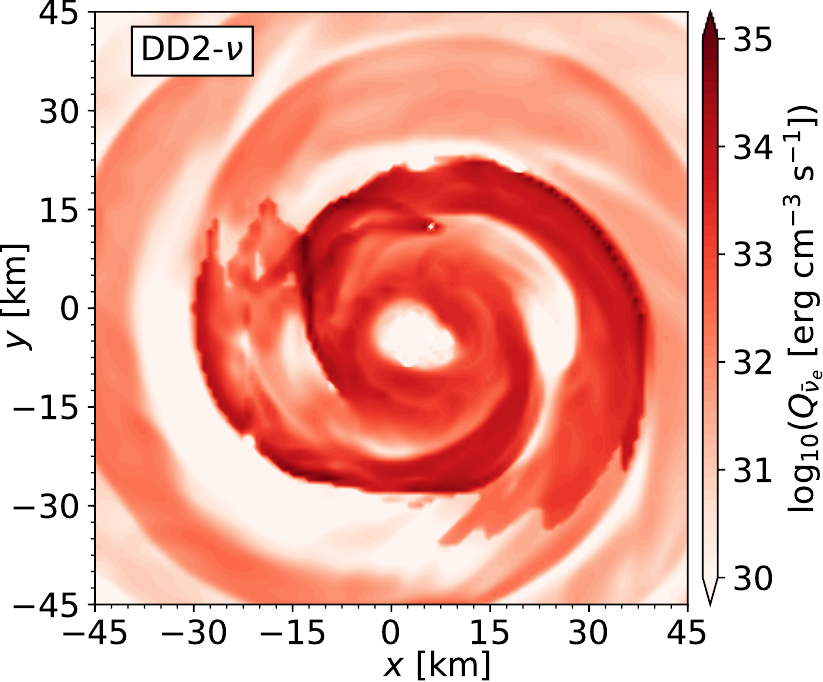}
\caption{Effective emissivities $10~{\rm ms}$ after the merger in the $x$-$y$ plane of electron neutrinos (\mbox{{\textbf{left} panels}}) and electron antineutrinos ({\textbf{right} panels}) for the SFHo-$\nu$ ({\textbf{upper} panels}) and DD2-$\nu$ ({\textbf{lower} panels}) runs. Here, we notice that the emissions are small at the densest portion of the core, concentrate at the hot parts of the disk/spiral arms, and decrease towards the outer regions of \mbox{the disk.}}
\label{fig:13}
\end{figure}

In Figure~\ref{fig:13}, we present the electron-flavored neutrino emissivities in the $x$-$y$ plane for the SFHo-$\nu$ and DD2-$\nu$ simulations. It is notable that the inner core has the smallest emissivities, which is expected from its typically high optical depth. In addition, both electron neutrinos and antineutrinos are emitted at very close rates, which amounts to a near-conservation of the electron fraction shortly after the merger. In addition, one notices that at specific regions of the remnant (e.g., in disk interfaces and outer portions of the spiral arms), the electron antineutrinos have emissivities that may be one order of magnitude higher than those of the electron neutrinos, which explains the luminosity dominance of electron antineutrinos over electron neutrinos in Figure~\ref{fig:9}. Besides, in the regions where the electron antineutrinos' emissivity is larger than the electron neutrinos' emissivity, the matter is then leptonized as visible in the lower panels of Figure~\ref{fig:7}, referring to the NLS runs. On the contrary, in regions such as the outer disk, where the electron neutrinos emissivity is greater than that of the electron antineutrinos, the matter undergoes deleptonization.
\vspace{-6pt}

\subsection{Long Inspiral Simulations} \label{sec:5.3}

Our long simulations cover $\approx$ 14 orbits before the merger and are performed with four different resolutions (see Table~\ref{tab:2}; for a comparison, see Appendix~\ref{app:A} for a convergence study). The quasi-circular orbit has an eccentricity of $\approx$ $7\times 10^{-4}$ after applying an eccentricity reduction procedure~\cite{Kyutoku:2014yba,Dietrich:2015pxa} to the ID. In the next sections, unless stated otherwise, we use the \mbox{R4 resolution}. Our ID corresponds to an isentropic configuration with constant entropy per baryon equal to $s \sim 1~k_B$ and $\beta$-equilibrium.

\subsubsection{Post-Merger Stage}

The post-merger evolution is marked by a significant increase of the central rest-mass density over the simulations time, as depicted in Figure~\ref{fig:6}. This can be interpreted as a consequence of the phase-transition leading to the appearance of $\Lambda$ hyperons at high densities, which, in turn, reduces pressure support and softens the EoS. This is in accordance with the results of \cite{Radice:2016rys}, although their results refer to cold isothermal ID with the BHB$\Lambda \phi$ EoS.

For a better understanding of the NLS on the post-merger evolution, we show in Figure~\ref{fig:12} snapshots of the rest-mass density, temperature, and electron fraction of our runs in the $x$-$y$~plane 10~ms after the merger for two cases, with and without the NLS. In the upper panels, we note a remnant comprised of a massive core surrounded by a disk, which extends further for the NLS runs. In the middle panels, we see that the inner core is substantially colder than the interface between the core and the surrounding disk, which is then thermally supported by higher pressure exerted by the hot material. The higher core temperature compared to that of the short inspiral simulations is reminiscent of the isentropic initial condition, by which $T_{\rm core}(t = 0) \approx 25~{\rm MeV}$, and it is not significantly altered during the inspiral and coalescence. In fact, the small variation of the core temperature for different EoSs is due to the generally weak dependency of the pressure on the temperature for high densities. This remains true for the NLS runs because the core is also optically thick; hence, neutrinos do not provide sufficient cooling within the simulation timespan. Similarly to the short inspiral runs, in the lower panels, we see that the remnants are neutron rich, with $Y_e \lesssim 0.1$ at the core and the disk when neutrinos are not considered. With the NLS, we observe a slight deleptonization of the core, an increase in the electron fraction at the heated arms, and an overall strong deleptonization of the outer disk regions, which can also be interpreted in light of the neutrino emissions' geometry (see Section~\ref{sec:5.3.2}).

\subsubsection{Neutrino Emissions}\label{sec:5.3.2}

To the best of our knowledge, there are no previous studies treating the features of an NLS implementation for an isentropic ID of the BHB$\Lambda \phi$ EoS. Therefore, in this section, we present and discuss our findings, focusing our comparisons on the short inspiral SFHo-$\nu$ and DD2-$\nu$ runs.

\begin{figure}[H]

\begin{minipage}{0.5\textwidth}
  
  \includegraphics[width=\linewidth]{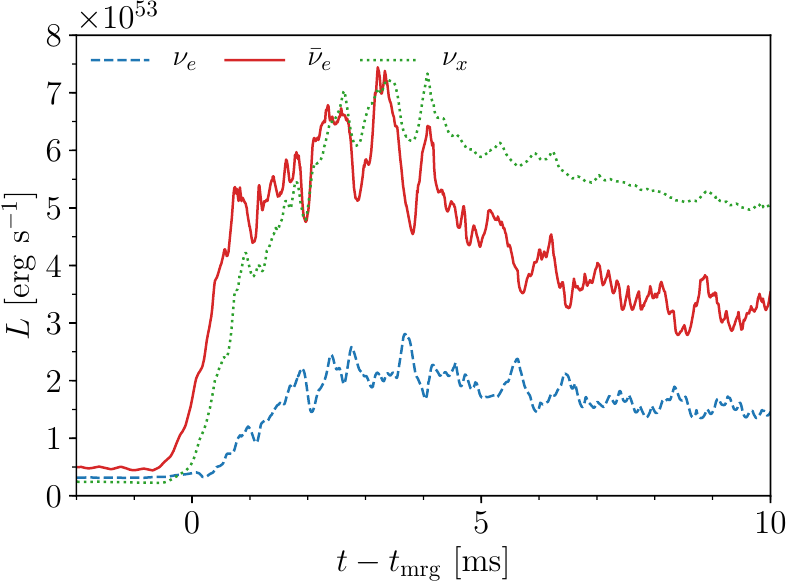}
\end{minipage}
\caption{Neutrino luminosity evolution for the BHB$\Lambda \phi$-$\nu$ simulation. Similarly to the short inspiral simulations, we have peak luminosities at $\sim$ 2--3 {ms} after the merger, followed by a decrease towards the end of the simulation.}
\label{fig:14}
\end{figure}

We start by presenting the luminosity evolution of the three neutrino species in Figure~\ref{fig:14}. One notices that during the late inspiral, the luminosities are greater than that of Figure~\ref{fig:9} and may be interpreted as a consequence of the higher temperatures within the coalescing NSs with isentropic thermal profile. The appearance of luminosity peaks at $t - t_{\rm mrg} \sim$ 2--3~{ms} followed by a decrease in individual luminosities is similar to the behavior of Figure~\ref{fig:9}, suggesting that this emission structure is an effect of our NLS implementation rather than an EoS-dependent feature. Additionally, we note that the peak luminosities for the electron antineutrinos and heavy lepton neutrinos reach higher values than those of the short inspiral runs. However, it is difficult to determine if this is caused by the thermal profile, the employed EOS, or by the higher masses of the merging NSs. By the end of the simulation, $L_{\bar\nu_e}$ dominates $L_{\nu_e}$ by a factor of $\sim$ $2.2$, and hence within the range obtained for the cold, low-mass SFHo and DD2 BNSs.
\begin{figure}[H]

\begin{minipage}{0.4\textwidth}
  
  \includegraphics[width=\linewidth]{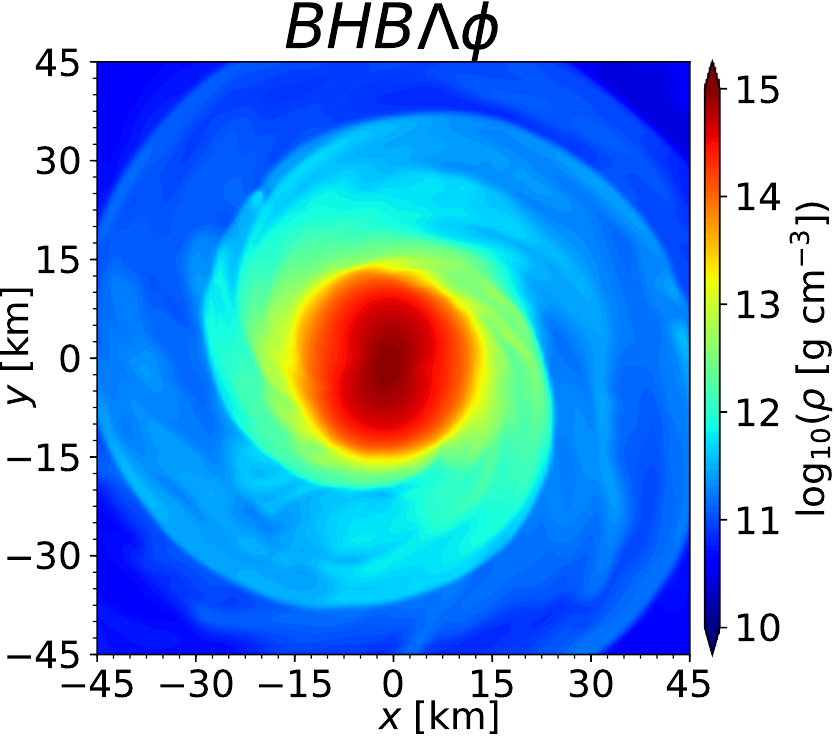}
\end{minipage}\hspace{0.5cm}
\begin{minipage}{0.4\textwidth}
  
  \includegraphics[width=\linewidth]{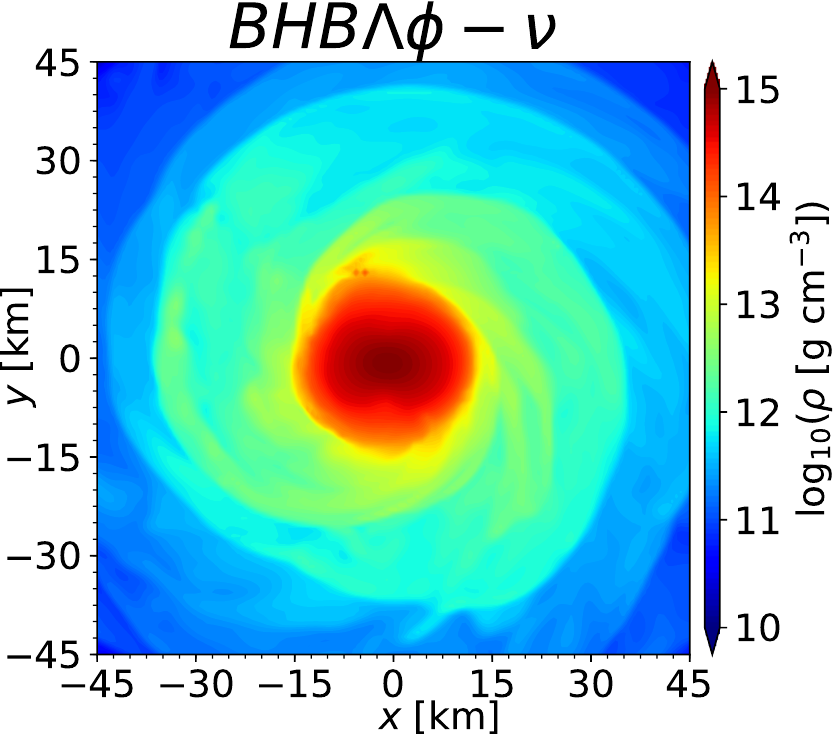}
\end{minipage}\\
\begin{minipage}{0.4\textwidth}
  
  \includegraphics[width=\linewidth]{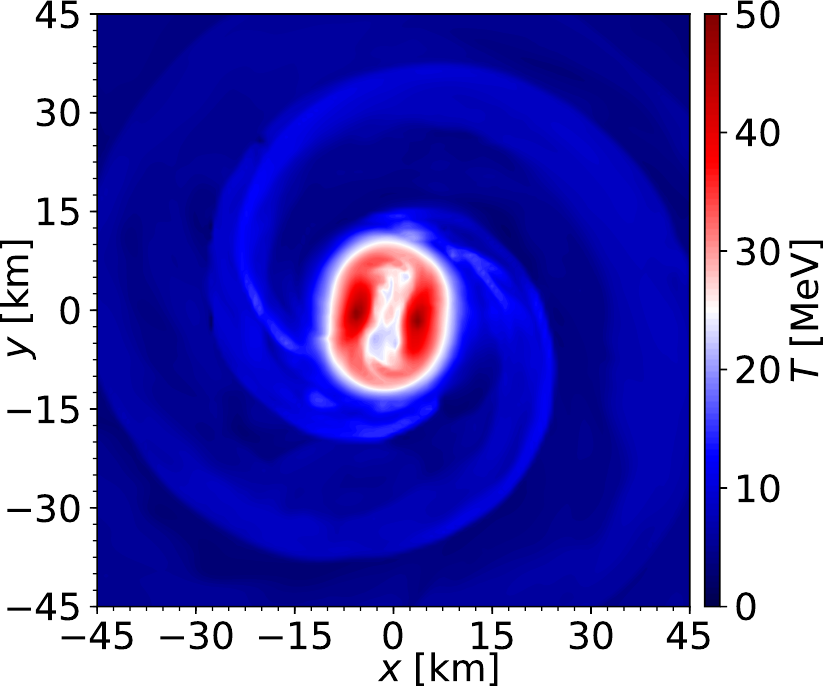}
\end{minipage}\hspace{0.5cm}
\begin{minipage}{0.4\textwidth}
  
  \includegraphics[width=\linewidth]{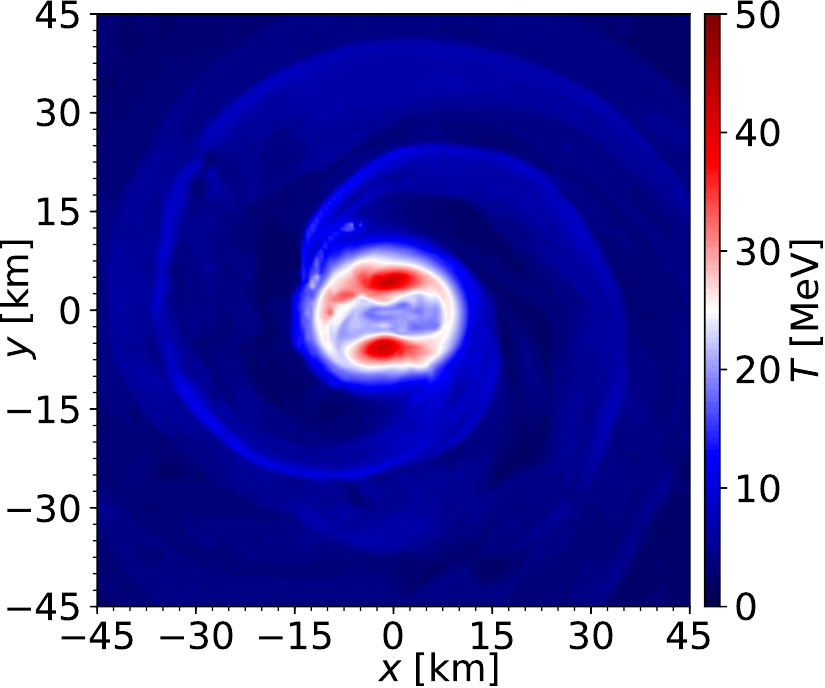}
\end{minipage}\\
\begin{minipage}{0.4\textwidth}
  
  \includegraphics[width=\linewidth]{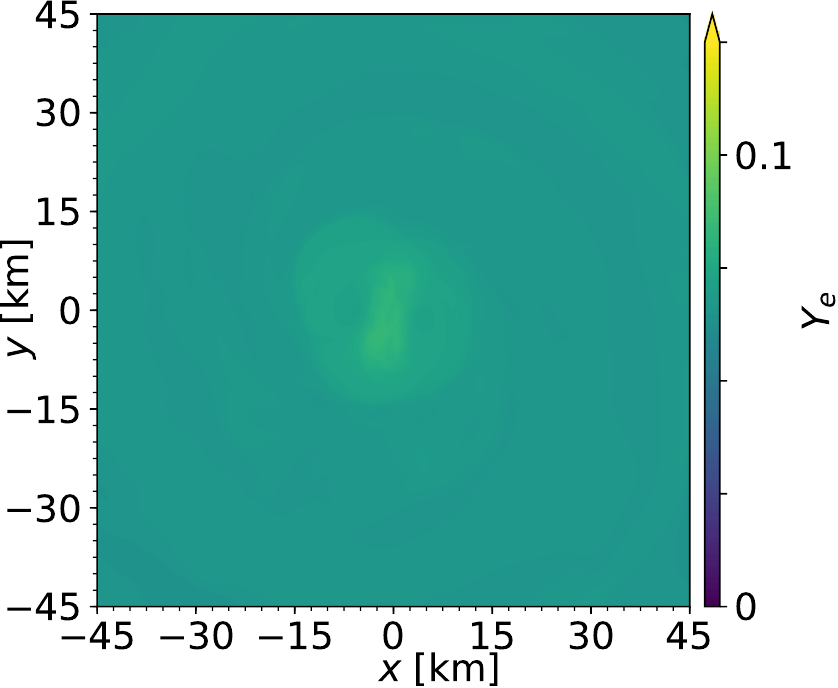}
\end{minipage}\hspace{0.5cm}
\begin{minipage}{0.4\textwidth}
  
  \includegraphics[width=\linewidth]{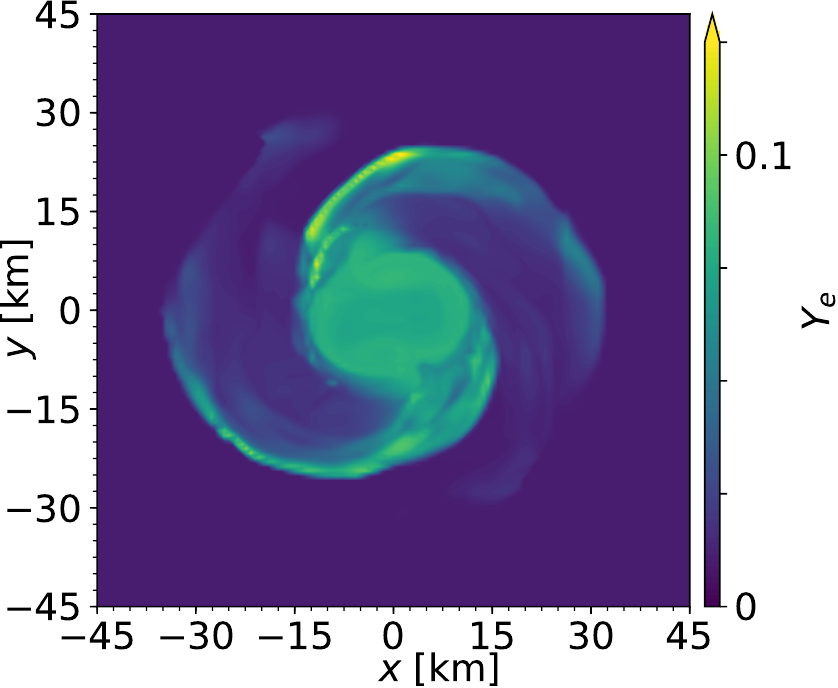}
\end{minipage}
\caption{Snapshots of hydrodynamical quantities of the simulations using BHB$\Lambda \phi$ and BHB$\Lambda \phi$-$\nu$ $10~{\rm ms}$ after the merger on the $x$-$y$ plane. From top to bottom, we have the rest-mass density, temperature, and electron fraction.}
\label{fig:12}
\end{figure}\vspace{-6pt}

\begin{figure}[H]

\begin{minipage}{0.45\textwidth}
  
  \includegraphics[width=\linewidth]{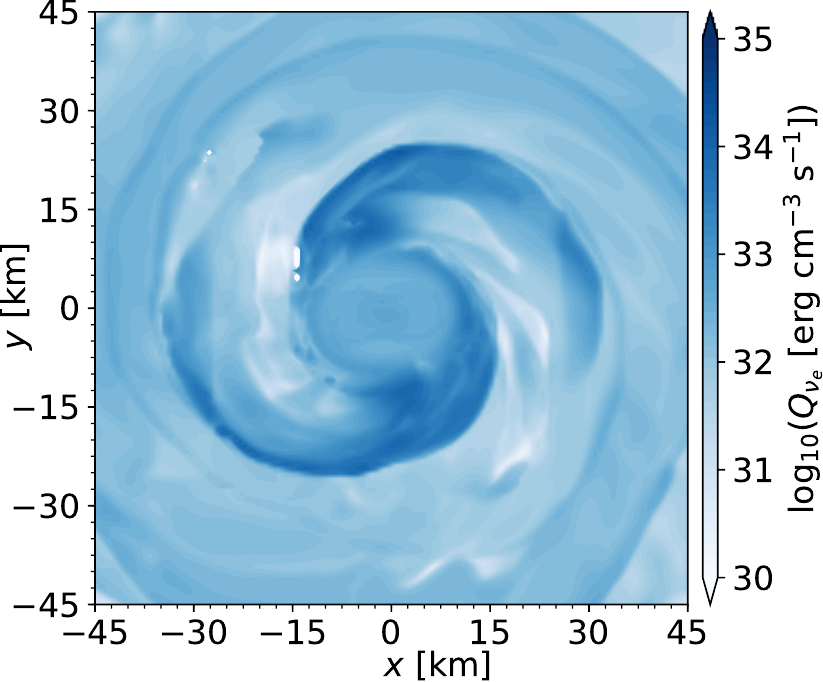}
\end{minipage}\hspace{0.5cm}
\begin{minipage}{0.45\textwidth}
  
  \includegraphics[width=\linewidth]{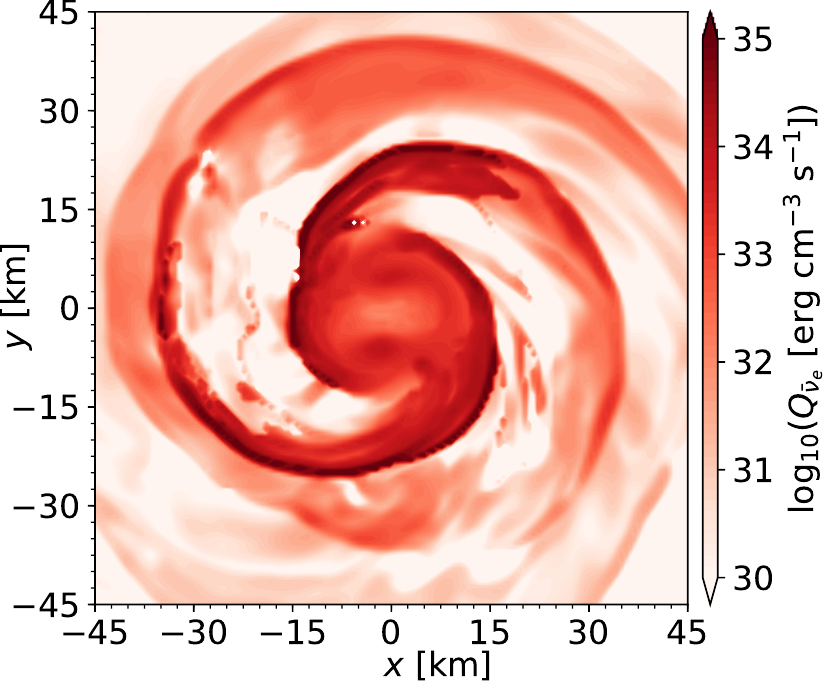}
\end{minipage}
\caption{Effective emissivities $10~{\rm ms}$ after the merger in the $x$-$y$ plane of electron neutrinos (\mbox{{\textbf{left} panels}}) and electron antineutrinos ({\textbf{right} panels}) for the BHB$\Lambda \phi$-$\nu$ run. Differently than for the cold ID case, we notice that the emissivities at the core are greater as a consequence of the isentropic thermal profile, which produces temperatures of tens of MeV within the NSs. The more potent emissions are found in the outer core and along the spiral arms, mostly in the inner disk region.}
\label{fig:15}
\end{figure}

Likewise, the electron fraction profile of the lower right panel of Figure~\ref{fig:12} is explained by the emission geometry depicted in Figure~\ref{fig:15}. It is interesting that the emissivities at the core of the remnant are greater than $10^{32}~{\rm erg~cm^{-3}~s^{-1}}$, which is more than two orders of magnitude higher than the counterparts of cold IDs (<$10^{30}~{\rm erg~cm^{-3}~s^{-1}}$). This is due to the core temperatures $T_{\rm core} \sim$ 20--30~{MeV}, which are reminiscent of the isentropic thermal profile of the ID. Finally, the leptonized portions of the remnant correspond to the regions where electron antineutrinos are more abundantly produced than electron neutrinos, namely at the core and the spiral arms at the inner portion of the disk. Conversely, in the remaining regions where $Q_{\nu_e} > Q_{\bar\nu_e}$, the fluid is deleptonized.

\section{\label{sec:6} Conclusions}

In this work, we presented first results of the extended infrastructure of the BAM code, with which we performed dynamical evolution of matter described by nuclear-theory-based three-parameter EoSs and neutrinos effects via an NLS. 

As a testbed of our new code framework, we simulated radially unstable TOV stars in full GR without the NLS in order to validate our GRHD implementation. This leads to a stable evolution within our simulations' timespan of {$\sim$ $5~{\rm ms}$}. A key observation regarding these tests is that the use of the high-order reconstruction scheme WENOZ introduces sufficiently small numerical viscosity so that radial stability is achieved by ejection of outer material layers {at later times}. We repeated these simulations with the same matter and grid configurations, but employing our NLS. We found that the NS cools and deleptonizes, and ultimately undergoes gravitational collapse in less than {$3~{\rm ms}$}.

Finally, we presented a set of BNS simulations using nuclear-theory EoSs and the NLS. In order to assess the capability of our code to handle distinct microphysical descriptions, we chose EoSs ranging from both ends of compactness (e.g., the soft SFHo and the stiff DD2), and including hyperons as of the BHB$\Lambda \phi$. We restricted our study to equal-mass, non-rotating systems.

Our short inspiral runs were performed with the DD2 and SFHo EoSs, with and without NLS, for the purpose of comparing results with the literature. Following \cite{Foucart:2015gaa}, we consider NSs with a gravitational mass in isolation of $M = 1.2~M_\odot$, initially at $\beta$-equilibrium and constant temperature $T = 0.1~{\rm MeV}$. Comparing our results to those of the aforementioned reference, we found good agreement in the formation of bar-like remnants that are stable during the simulations' timespan, surrounded by thermally supported, thick, and dense disks with pronounced spiral arms. The density and temperatures on the equatorial plane by the end of our simulations are also very {similar} to those of \cite{Foucart:2015gaa}. However, our electron fraction on the remnant is overall smaller, especially at the outer disk. This seems to be caused by our NLS implementation; in particular, we may be underestimating optical depths as of Equation~\eqref{eq:opt-est}, and hence predicting larger effective emission rates. This {may be a factor to explain} the peaks of total luminosity presented in Figure~\ref{fig:9}, which are larger than the peaks of \cite{Foucart:2015gaa} by a factor of two, {but further investigation is needed to single out the role of our optical depths prescription.} It is worth pointing out, though, that our results are consistent in the sense that we indeed find deleptonized matter in regions where the electron neutrinos emissivities are larger than those of the electron antineutrinos, and likewise the leptonized portions of the remnant coincide with those regions at which $Q_{\bar\nu_e} > Q_{\nu_e}$. In addition, the GWs spectra of our simulations largely agree with those of \cite{Foucart:2015gaa}, reproducing similar properties.

We also performed long inspiral simulations for various grid resolutions using the BHB$\Lambda \phi$ EoS, with and without the NLS, initially with constant entropy per baryon $s \sim 1~{\rm k_B}$ and in $\beta$-equilibrium. We found that the maximum rest-mass densities evolution during the post-merger stage is very similar to that of \cite{Radice:2016rys}, although they use a cold ID for this EoS, which suggests that isentropy (and consequently, the initial thermal profile) is not that relevant for the evolution of the densest portion of the remnant. The core temperature is not significantly altered during the inspiral and coalescence, remaining at $T_{\rm core} \approx 25~{\rm MeV}$ by the end of the simulations, which is reminiscent of the initial isentropic thermal profile. Similarly to the short inspiral simulations, the remnant was mostly deleptonized by the end of our simulations.

Overall, our implementations allowed long-term stable, constraint-satisfying evolutions, with a performance comparable to the previous version of the BAM code (see Appendix~\ref{app:A}), despite the increase in complexity and realism encompassed by our new framework. As of Figure~\ref{fig:A2}, we found that the GW phase difference of $(2,2)$ decreases with increasing resolution for the BHB$\Lambda \phi$ run, but in the case of BHB$\Lambda \phi$-$\nu$, there is no significant improvement by the increase of numerical resolution, which we will investigate further in the future. {Likewise, in future works, we intend to enhance our scheme by including muons in our GRHD formalism and related neutrinos processes, such as muons-driven Urca and modified Urca reactions, which might be important in BNS mergers~\cite{Alford:2018lhf}.}

\vspace{6pt} 



\authorcontributions{Conceptualization, T.D., M.U., and H.G.; methodology, T.D., M.U., and H.G.; software, T.D. and H.G.; validation, H.G.; formal analysis, T.D., M.U., H.G., and F.S.; investigation, T.D, M.U., and H.G.; resources, T.D.; data curation, T.D. and H.G.; writing---original draft preparation, T.D., M.U., H.G., and F.S.; writing---review and editing, T.D., M.U., H.G., and F.S.; visualization, H.G.; supervision, T.D. and M.U.; project administration, T.D. and M.U.; funding acquisition, T.D., M.U., and H.G. All authors have read and agreed to the published version of the manuscript.}

\funding{This research was funded by FAPESP grant number 2019/26287-0.
The simulations were performed on the national supercomputer HPE Apollo Hawk at the High-Performance Computing (HPC) Center Stuttgart (HLRS) under the grant number GWanalysis/44189, 
on the GCS Supercomputer SuperMUC at Leibniz Supercomputing Centre (LRZ) (project pn29ba), 
and the HPC systems Lise/Emmy of the North German Supercomputing Alliance (HLRN) (project bbp00049). }

\institutionalreview{Not applicable.} 

\informedconsent{Not applicable.} 


\dataavailability{The data presented in this study are available on request from the corresponding author. The data are not publicly available due to its large size.} 

\acknowledgments{H.G.  and M.U.  thank FAPESP for financial support. M.U.  thanks CAPES (through the Coordenação de Aperfeiçoamento de Pessoal de Nível Superior, Brasil (CAPES), process number: 88887.571346/2020-00) for financial support to visit the University of Potsdam during the final stages of this project, and thanks the University of Potsdam for its hospitality. }

\conflictsofinterest{The authors declare no conflict of interest.}

\appendixtitles{yes} 
\appendixstart
\appendix
\section{\label{app:A} Convergence of the Code}

\begin{figure}[H]

\includegraphics[width=0.49\linewidth]{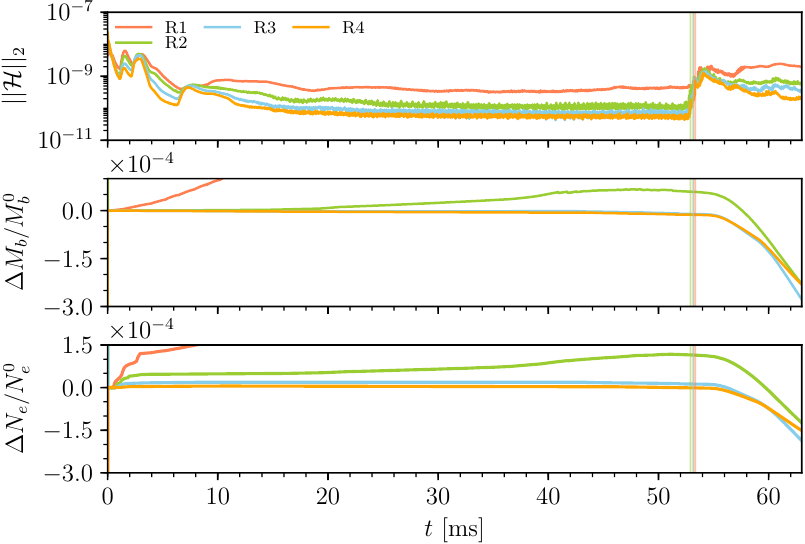}
\includegraphics[width=0.49\linewidth]{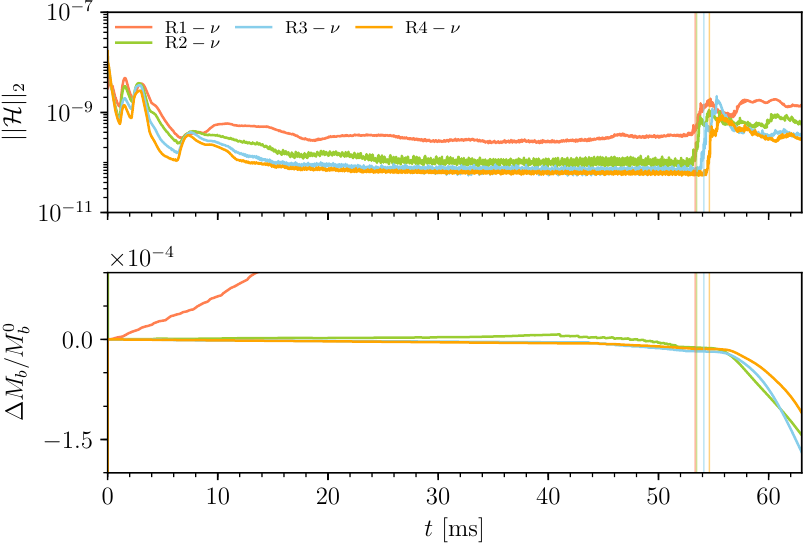}
\caption{\textls[-10]{Constraint evolution and conserved quantities. {\textbf{Left} panel:} BHB$\Lambda \phi$. {\textbf{Right} panel:} \mbox{BHB$\Lambda \phi$-$\nu$.}} We show the $L2$ norm of the Hamiltonian constraint $||\mathcal{H}||_2$, the baryonic mass variation $\Delta M_b/M_b^0$, and the electrons number variation $\Delta N_e/N_e^0$ for the BHB$\Lambda \phi$ (left panel) for all four resolutions of Table~\ref{tab:2}. We observe good behavior of the Hamiltonian constraint and conserved quantities with increasing resolution in both cases. Physical quantities were extracted from the grid level $l=1$, and the vertical lines mark the merger for each resolution.}
\label{fig:A1}
\end{figure}

We present in the left panel of Figure~\ref{fig:A1} the evolution of the $L2$ norm of the Hamiltonian constraint $||\mathcal{H}||_2$, the variation of the baryonic mass $\Delta M_b$ (with respect to its initial value $M_b^0$), and the variation of the number of electrons $\Delta N_e$ (with respect to its initial value $N_e^0$) for the four resolutions employed for the BHB$\Lambda \phi$ EoS simulations (see Table~\ref{tab:2}). In the right panel, the same physical quantities are depicted for the BHB$\Lambda \phi$-$\nu$ case, but not for the number of electrons, which is not conserved in the implemented scheme. The behavior of the constraints and the conservation of the physical quantities improves with increasing grid resolution for both cases. For the BHB$\Lambda \phi$ case, we see that for resolution R$2$ up to resolution R$4$, the Hamiltonian constraint begins at values of order $\mathcal{O}(10^{-8})$ (mainly due to the loading of the initial data), but rapidly decreases to order $\mathcal{O}(10^{-10})$ and stays at this level during the majority of the time, only increasing to a stable $\leq \mathcal{O}(10^{-9})$ during the post-merger. The baryonic mass is conserved to order $\mathcal{O}(10^{-5})$ for resolutions R$3$ and R$4$, which is comparable to the results obtained when pwps are used. The conservation of the number of electrons, which was previously not included in BAM, presents small violations of order $\mathcal{O}(10^{-4})$ for resolutions R$2$,~R$3$, and R$4$. The increase in the R$2$ electron number conservation violation after $t \sim 40~{\rm ms}$ is absent in runs R$3$ and R$4$, which suggests that this is indeed a {resolution}-dependent effect. For the BHB$\Lambda \phi$-$\nu$ case, the values of the Hamiltonian constraint stay at order $\mathcal{O}(10^{-10})$ for resolutions R$3$ and R$4$, and increase to a stable $\mathcal{O}(10^{-9})$ during the post-merger. The baryons conservation is violated to less than $\mathcal{O}(10^{-4})$ for the majority of the run, and the adoption of the NLS seems to improve the behavior of the R2 resolution when compared to the BHB$\Lambda \phi$ case. In both cases, we found that after the merger, the conserved quantities accuracy is less efficient, mainly due to the artificial atmosphere scheme used in BAM, in which ejected material with low density is treated as atmosphere. 

Overall, the results presented in Figure~\ref{fig:A1} have constraint violations of the same order when compared to simulations using the previous version of BAM~\cite{Bernuzzi:2016pie}, which relied on a simpler description of the matter using pwp EoSs.

\begin{figure}[H]

\includegraphics[width=0.49\textwidth]{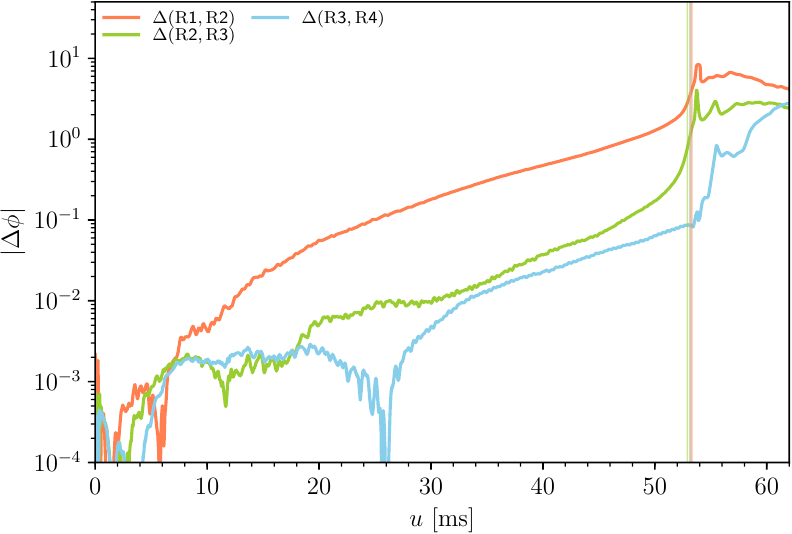}
\includegraphics[width=0.49\textwidth]{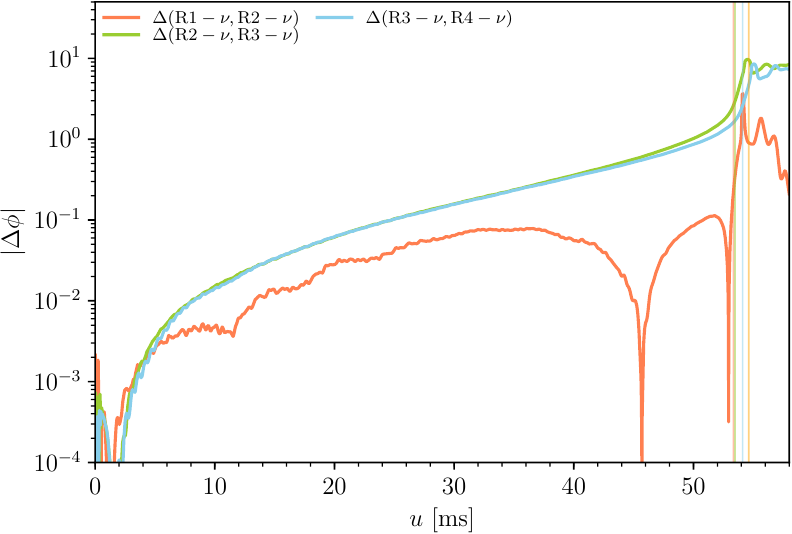}
\caption{Phase difference of the $(2,2)$ mode of the GW signal between different resolutions with respect to the retarded time $u$. The BHB$\Lambda \phi$ case is presented on the \textbf{left} panel and the BHB$\Lambda \phi$-$\nu$ case is presented on the \textbf{right} panel. Solid vertical lines mark the merger for each resolution.}
\label{fig:A2}
\end{figure}

In Figure~\ref{fig:A2}, we present the convergence plots of the GW $(2,2)$ mode phase obtained at the outermost extracted radius of the computational domain ($\approx$886~km). The BHBcase $\Lambda \phi$ is presented in the left, and the BHB$\Lambda \phi$-$\nu$ case in the right panel. The difference between the phases of the $R1$ and $R2$ resolutions, $|\Delta(R1,R2)|$, is depicted for completeness because resolution R1 clearly does not conserve the physical quantities along the evolution (see Figure~\ref{fig:A1}). Using the other numerical resolutions, no clear convergence order can be estimated from the plots using $|\Delta(R2,R3)|$ and $|\Delta(R3,R4)|$, and the small difference between these quantities suggests that the increasing resolution does not significantly improve the results (see, for example, the right panel of Figure~\ref{fig:A1}, where resolutions $R2$, $R3$, and$ R4$ give practically the same results). In the future, we plan to employ the higher-order method outlined in~\cite{Bernuzzi:2016pie} or an entropy-limited scheme such as that in \cite{Doulis:2022vkx}.

\begin{adjustwidth}{-\extralength}{0cm}
\reftitle{References}

\end{adjustwidth}
\end{document}